\documentclass[preprint,aps,pra,superscriptaddress,floatfix,nofootinbib]{revtex4-1}
\usepackage{array}
\usepackage{relsize}
\usepackage{placeins}
\usepackage{amsmath}
\usepackage{mathrsfs}
\usepackage{amssymb}
\usepackage{mathtools}
\usepackage{graphicx}
\usepackage{color}
\usepackage{hyperref}
\usepackage{subfigure}
\usepackage[normalem]{ulem}
\usepackage{float}

\usepackage[final]{changes}  

\newcommand{\be}{\begin{equation}}
\newcommand{\ee}{\end{equation}} 
\newcommand{\lb}{\label}
\newcommand{\OL}{\overline}

\newcommand{\const}{({\rm const.})}

\newcommand{\ba}{{\bf a}}

\newcommand{\bF}{{\bf F}}

\newcommand{\bk}{{\bf k}}

\newcommand{\br}{{\bf r}}
\newcommand{\bs}{{\bf s}}
\newcommand{\bu}{{\bf u}}

\newcommand{\bv}{{\bf v}}
\newcommand{\bx}{{\bf x}}

\newcommand{\bS}{{\bf S}}

\newcommand{\bOmega}{{\mbox{\boldmath $\Omega$}}}
\newcommand{\bomega}{{\mbox{\boldmath $\omega$}}}

\newcommand{\grad}{{\mbox{\boldmath $\nabla$}}}
\newcommand{\bdot}{{\mbox{\boldmath $\cdot$}}}

\newcommand{\btimes}{{\mbox{\boldmath $\times$}}}

\begin{document}

\title{\textbf {Effective Drift Velocity from Turbulent Transport by Vorticity}}
\author{Hussein Aluie}\email{hussein@rochester.edu}
\affiliation{Department of Mechanical Engineering, University of Rochester, Rochester, New York}
\affiliation{Laboratory for Laser Energetics, University of Rochester, Rochester, New York}
\author{Shikhar Rai}
\affiliation{Department of Mechanical Engineering, University of Rochester, Rochester, New York}
\author{Hao Yin}
\affiliation{Department of Mechanical Engineering, University of Rochester, Rochester, New York}
\author{Aarne Lees}
\affiliation{Department of Mechanical Engineering, University of Rochester, Rochester, New York}
\affiliation{Laboratory for Laser Energetics, University of Rochester, Rochester, New York}
\author{Dongxiao Zhao}
\affiliation{Department of Mechanical Engineering, University of Rochester, Rochester, New York}
\author{Stephen M. Griffies}
\affiliation{NOAA/Geophysical Fluid Dynamics Laboratory, Princeton, New Jersey}
\affiliation{Princeton University Program in Atmospheric and Oceanic Sciences, Princeton, New Jersey}
\author{Alistair Adcroft}
\affiliation{NOAA/Geophysical Fluid Dynamics Laboratory, Princeton, New Jersey}
\affiliation{Princeton University Program in Atmospheric and Oceanic Sciences, Princeton, New Jersey}
\author{Jessica K. Shang}
\affiliation{Department of Mechanical Engineering, University of Rochester, Rochester, New York}
\affiliation{Laboratory for Laser Energetics, University of Rochester, Rochester, New York}

\begin{abstract}
We highlight the differing roles of vorticity and strain in the transport of coarse-grained scalars at length-scales larger than $\ell$ by smaller scale (subscale) turbulence.
We use the first term in a multiscale gradient expansion due to Eyink \cite{Eyink06a}, which exhibits excellent correlation with the exact subscale physics when the partitioning length $\ell$ is any scale smaller than that of the spectral peak. We show that unlike subscale strain, which acts as an anisotropic diffusion/anti-diffusion tensor, subscale vorticity's contribution is solely a conservative advection of coarse-grained quantities by an eddy-induced non-divergent velocity, $\bv_*$, that is proportional to the curl of vorticity. Therefore, material (Lagrangian) advection of coarse-grained quantities is accomplished not by the coarse-grained flow velocity, $\OL\bu_\ell$, but by the effective velocity, $\OL\bu_\ell+\bv_*$, the physics of which may improve commonly used LES models.
\end{abstract}

\maketitle

%
 
\section{Introduction}

Basic considerations from fluid dynamics (e.g. \cite{kundu2015fluid}) indicate that the distance between particles in a laminar flow is determined by the strain. Vorticity merely imparts a rotation on their separation vector $\br$ without affecting its magnitude. This behavior can be seen by considering the velocity, $\bu$, difference between particles $P$ and $Q$ at positions $\bx$ and $\bx+\br$, respectively, 
\begin{eqnarray}
\bu_{Q}-\bu_{P}= \delta\bu = \bu(\bx+\br)-\bu(\bx) = \br\bdot\grad\bu{\big|}_{\bx}+ \dots~,
\end{eqnarray}
where a Taylor-series expansion is justified for short distances $|\br|$ over which the flow is sufficiently smooth. In the Lagrangian frame of $P$ at $\bx$, the separation from $Q$ evolves as 
\begin{eqnarray}
\frac{D \br}{Dt} = \delta\bu = \br\bdot\bS + \underbrace{~\br\bdot\bOmega~}_{\frac{1}{2}\bomega\btimes\br},
\lb{eq:DrDt}\end{eqnarray}
where the velocity gradient tensor, $\grad\bu = \bS + \bOmega$, has been  decomposed into the symmetric strain rate tensor $\bS = [\grad\bu + (\grad\bu)^{T}]/2$ and the antisymmetric vorticity tensor $\bOmega = [\grad\bu - (\grad\bu)^{T}]/2 = -\frac{1}{2} \epsilon_{ijk} \omega_k$. Here, $\bomega=\grad\btimes\bu$ is vorticity and $\epsilon_{ijk}$ is the Levi-Civita symbol. Taking an inner product of eq.~\eqref{eq:DrDt} with $\br$, 
\begin{eqnarray}
\frac{1}{2}\frac{D |\br|^2}{dt} = \br\bdot\bS\bdot\br~, 
\end{eqnarray}
shows that the distance is determined by the strain. Vorticity in eq.~\eqref{eq:DrDt} only acts to rotate $\br$ without changing its magnitude.

\begin{figure}[ht]
    \centering
    \includegraphics[width=0.75\textwidth]{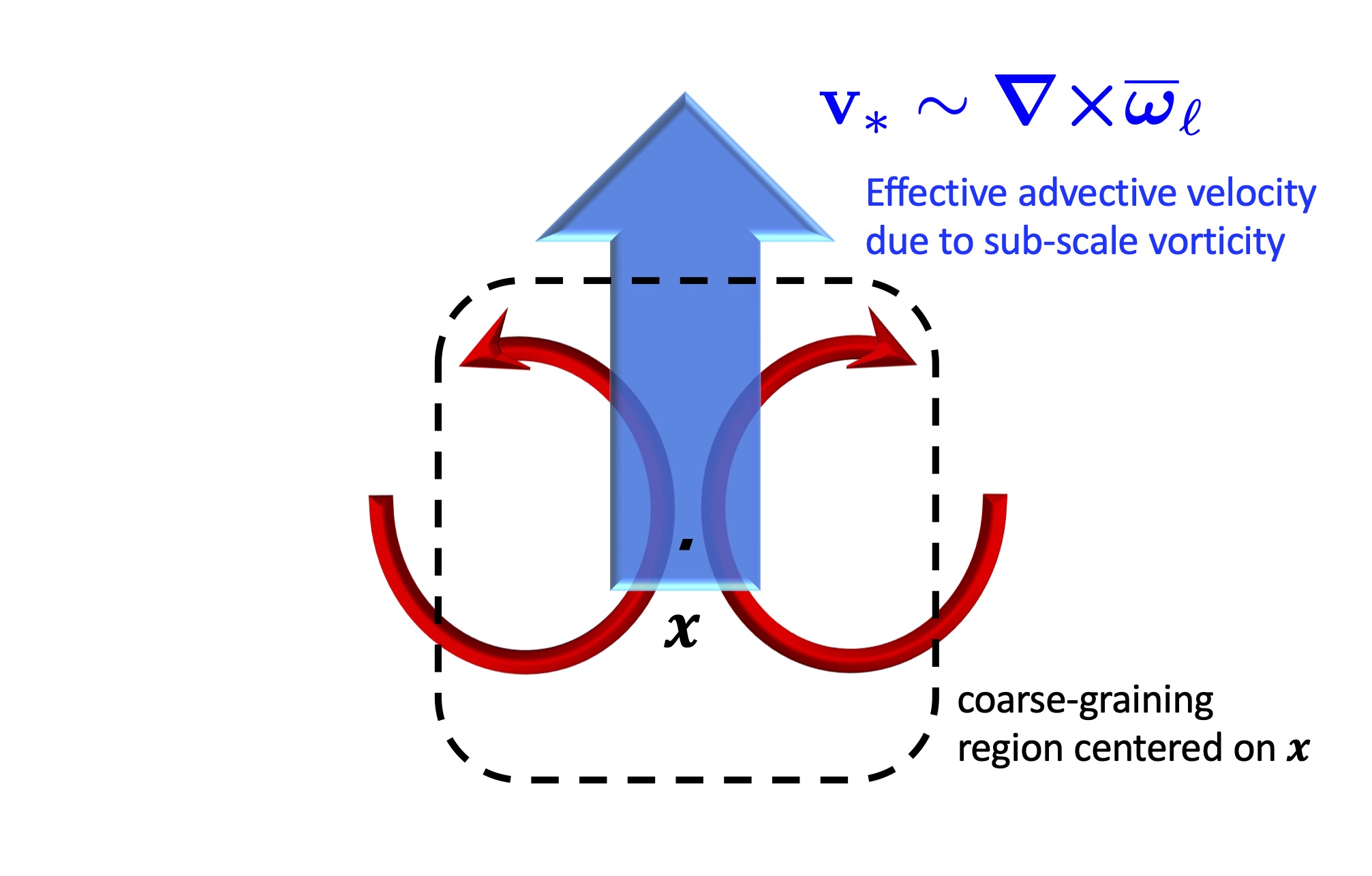}
    \caption{Schematic of how eddy-induced advection by a non-divergent velocity, $\bv_*\sim\grad\btimes\OL{\bomega}_\ell$, (blue arrow) from eq.~\eqref{eq:EffectVel} can arise due to subscale vorticity (red). In this example, two symmetric but counter-rotating eddies (red) are of a size smaller than that of the coarse-graining box (dashed). At length-scales larger than that of the box, $\ell$, which may represent a grid cell in a simulation or the resolution limit of an observation, these eddies exert an eddy-induced advection with velocity $\bv_*$ proportional to the curl of vorticity, depicted by the blue arrow at location $\bx$.
    }
    \label{fig:EffectVort}
\end{figure}

These considerations hinge on the critical assumption that the flow is sufficiently smooth over separations $\br$, which is patently invalid in a turbulent flow for $\br$ at inertial scales \cite{Pope00}. However, a version of this story survives thanks to the property of scale-locality, which justifies an expansion in scale as we shall discuss below. The main result of this paper is eq.~\eqref{eq:EffectVel}, which is an expression for the eddy-induced advection velocity $\bv_*$ at length-scales larger than $\ell$, which may be the size of grid cells in a simulation. The non-divergent velocity, $\bv_*$, arises from vortical motions associated with subscale nonlinear interactions.
Fig.~\ref{fig:EffectVort} captures the essential insight behind eq.~\eqref{eq:EffectVel}. 

In section~\ref{sec:Filtering}, we discuss subscale transport, Eyink's expansion in length-scale, and the leading order term. In section~\ref{sec:Empirical}, we present empirical support from (i) 3D direct numerical simulation (DNS) of a compressible turbulent flow and (ii) two-layer stacked shallow water simulation of geophysical fluid turbulence. Section~\ref{sec:MainResult} contains the main result, which expresses transport by subscale vorticity as a conservative advection by an effective drift velocity $\bv_*$ proportional to the curl of vorticity. Throughout this paper we attempt to make the presentation  accessible to readers from various backgrounds. 

\section{Multi-Scale Dynamics}\lb{sec:Filtering}
To analyze the dynamics of different scales in a flow, we use the coarse-graining approach,
which has proven to be a natural and versatile framework to understand and model scale interactions 
(e.g. \cite{MeneveauKatz00,Eyink05}). 
The approach is standard in partial differential equations
and distribution theory (e.g.  \cite{Strichartz03,Evans10}). 
It became common in Large Eddy Simulation (LES) modeling of turbulence thanks to the foundational works of Leonard \cite{Leonard74} and Germano \cite{Germano92}.

For any field $\ba(\bx)$, a coarse-grained or (low-pass) filtered version of this field, which contains spatial variations
at scales $>\ell$, is defined in $n$-dimensions as
\be
\OL \ba_\ell(\bx) = \int \mathrm{d}^n\br~ G_\ell(\bx-\br)\, \ba(\br).
\lb{eq:filtering}\ee
Here, $G_\ell$ is a convolution kernel derived from a
 ``mother'' kernel $G(\bs)$ (borrowing the term from wavelet analysis \cite{daubechies1992ten}). $G(\bs)$ is normalized, $\int \mathrm{d}^n \bs\, G(\bs) = 1$. It is an even function such that, $\int \bs\, G(\bs) \, \mathrm{d}^n \bs = 0$, which ensures that local averaging is symmetric.
$G(\bs)$ also has its main support (or variance) over a region of order unity in diameter, $\int \mathrm{d}^n \, \bs\,|\bs|^2\,G(\bs)=O(1)$. The dilated version of the kernel, 
\be
G_\ell(\br)= \ell^{-n} G(\br/\ell),
\lb{eq:dilatedKernel}\ee
inherits all those properties, except that its main support is over a region of diameter $\ell$. An example is the Top-Hat kernel in non-dimensional coordinates $\bs$,
\begin{eqnarray}
G(\bs)=
\begin{cases}
1, &\text{if $|s_i|<1/2$ for $i=1,\dots,n$},\\
 0,  &\text{otherwise},\\
\end{cases}
\lb{eq:TopHat_mother}
\end{eqnarray}
and its dilated version in dimensional coordinates $\br$,
\begin{eqnarray}
G_{\ell}(\br)=
\begin{cases}
\ell^{-n}, &\text{if $|r_i|<\ell/2$ for $i=1,\dots,n$},\\
 0,  &\text{otherwise}.\\
\end{cases}
\lb{eq:TopHat}\end{eqnarray}

The scale decomposition in eq. (\ref{eq:filtering}) is essentially a partitioning of scales for a field into large scales ($\gtrsim\ell$), captured by $\OL \ba_\ell$, and small scales ($\lesssim\ell$), captured by the residual 
\be
\ba'_\ell=\ba-\OL \ba_\ell.
\lb{eq:high-pass}\ee
In the remainder of this paper, we shall omit subscript $\ell$ from variables if there is no risk for confusion.

\subsection{Coarse-grained Transport}
Consider the transport equation of the concentration (per unit volume)  $C(\bx,t)$:
\be 
\partial_t C + \grad\bdot (C\,\bu) =  \grad\bdot[\rho \, \kappa \, \grad \phi].
\lb{eq:Advect}\ee

Here, $\rho$ is mass density, $\bu$ is the advecting flow velocity, which is possibly compressible. The scalar field $C=\rho \, \phi$, where $\phi$ is concentration per unit mass, can be either passive or active, with diffusivity $\kappa$, which can be due to the microphysics or an effective (turbulent) diffusivity used in a simulation.  Analysis of the incompressible momentum transport follows similar reasoning as below, with $C$ replaced by $\bu$ when $\rho=\const$, so we shall focus on $C$ for simplicity.

We shall call eq. \eqref{eq:Advect} the ``bare'' dynamics \cite{Aluie17,eyink2018review} as it governs flow in an experiment or natural system. In a coarse-resolution simulation, such as in implicit or explicit LES \cite{grinstein2007implicit,MeneveauKatz00}, or when taking spatially under-resolved measurements in an experiment, $\bu$ and $C$ are not resolved  down to the microphysical scales. The equation that governs the coarse-grained scalar $\OL{C}_\ell$ is obtained by filtering eq.~\eqref{eq:Advect} to get (e.g. \cite{higgins2004heat})
\begin{subequations}
\begin{align}
&\partial_t \OL{C}_\ell + \grad\bdot (\OL{\bu}_\ell \,\OL{C}_\ell) =  -\grad\bdot\OL\tau_\ell + \grad\bdot\OL{(\rho \, \kappa \grad \phi)}_\ell~,
\lb{eq:CoarseAdvect_a}\\
\mbox{where~~}&\OL\tau_\ell=\OL\tau_\ell(\bu,C) \equiv \OL{(\bu \, C)}_\ell - \OL{\bu}_\ell \, \OL{C}_\ell
\lb{eq:subscaleflux}
\end{align}
\lb{eq:CoarseAdvect}\end{subequations}
is the subscale flux of $\OL{C}_\ell$ \cite{Germano92}. Eq.\eqref{eq:CoarseAdvect} is exact, without any approximation, and describes the scalar transport at length-scales larger than $\ell$. The last term, $\grad\bdot\OL{(\rho \, \kappa \grad \phi)}_\ell$, was shown mathematically in \cite{Aluie13} to have an upper bound proportional to $\alpha_{\textrm{rms}} \,\ell^{-2} \,\phi_{\textrm{rms}}$ at every location in the flow and every time, where $\alpha=\rho \, \kappa$. Therefore, it is guaranteed to be negligible for sufficiently large $\ell$ or sufficiently small $\alpha$,
which was demonstrated numerically in \cite{ZhaoAluie18}. Therefore, $\grad\bdot\OL{(\rho \, \kappa \grad \phi)}_\ell$ can be dropped from eq.~\eqref{eq:CoarseAdvect} (but not from the bare eq.~\eqref{eq:Advect}).

An evident role of the subscale flux $\OL{\tau}_\ell(C,\bu)$ in eq.~\eqref{eq:CoarseAdvect} is accounting for the influence of unresolved scales $C'_\ell$ and $\bu'_\ell$ on the transport of $\OL{C}_\ell$. Another important role of $\OL{\tau}_\ell(C,\bu)$ in eq.~\eqref{eq:CoarseAdvect} that may not be as well-appreciated is balancing $\OL{\bu}_\ell\,\OL{C}_\ell$, which can interact with scales $<\ell$ that are approximately within the band $[\ell/2,\ell]$. This is analogous to the term $\rho\kappa\grad\phi$ balancing $\bu\,C$ at microphysical scales $\sim\ell_\kappa$, below which molecular diffusion dominates advection. 
To illustrate, assume that at a time $t$ the scalar and velocity fields are fully resolved, $\OL{C}_\ell = C$ and $\OL{\bu}_\ell = \bu$. Moreover, assume that scale $\ell$ is sufficiently large, $\ell\gg\ell_\kappa$, such that microphysical diffusivity is negligible and does not damp nonlinear interactions. That small scales are absent, $C'_\ell=0$ and $\bu'_\ell=0$, does not necessarily imply that the subscale flux $\OL{\tau}_\ell(C,\bu) = \OL{(\bu \, C)}_\ell - \bu \, C$ is zero unless there is sufficient scale separation between $\ell$ and the smallest scales of $C$ and $\bu$.
Therefore, the general transport of a coarse-grained scalar $\OL{C}_\ell$ is not described by the typical (or bare) transport equation~\eqref{eq:Advect} even when $C$ and $\bu$ are resolved ($C'_\ell=0$ and $\bu'_\ell=0$), because coarse-graining also alters the representation of their nonlinear interactions, which may be at scales smaller than $\ell$. 

From definition \eqref{eq:subscaleflux}, it is possible to  decompose $\OL{\tau}_\ell(C,\bu)$ into contributions that are solely due to resolved fields $\OL{C}_\ell$ and $\OL{\bu}_\ell$ (e.g. the Leonard term \cite{Leonard74,germano1986proposal}) and other contributions that involve unresolved fields $C'_\ell$ and $\bu'_\ell$. In the discussion below, we use the term `subscale' to refer to any contribution from $\OL{\tau}_\ell(C,\bu)$ because it arises from coarse-graining the dynamics at scale $\ell$. We shall see that the resolved fields make a dominant contribution to the subscale flux $\OL{\tau}_\ell(C,\bu)$ thanks to the property of ultraviolet scale-locality.

\subsection{Multiscale Expansion}\lb{sec:NLmodel}
The subscale term $\OL\tau_\ell(\bu,C)$ 
does not lend itself to a straightforward intuitive interpretation. To that end, we utilize the multiscale gradient expansion of Eyink \cite{Eyink06a}. This expansion is accomplished by decomposing a field 
\be \ba(\bx) = \sum_{p=0}^{\infty} \ba^{[p]}(\bx)
\ee
into a sum of band-passed fields \cite{Eyink06a}
\begin{subequations}
\begin{align}
\ba^{[p]} &= \OL\ba_{\ell_p} - \OL\ba_{\ell_{p-1}},~~~p=1,2,\dots,\\
\ba^{[0]} &= \OL\ba_{\ell_0} 
\end{align}
\end{subequations}
for a sequence of scales $\ell_p = \lambda^{-p}\ell_0$ with the non-dimensional number $\lambda > 1$, e.g. $\lambda=2$. $\ell_0$ can be a characteristic large scale such as the domain size. A similar decomposition can be done for the subscale flux,
\begin{subequations}
\begin{align}
&\OL\tau_\ell(\bu,C) = \sum_{p=0}^{\infty}\sum_{q=0}^{\infty}\tau_\ell^{[p,q]}(\bu,C),
\lb{eq:subscaleDecomp_a}\\
\mbox{with~~}&\tau_\ell^{[p,q]}(\bu,C) = \OL{(\bu^{[p]}C^{[q]})}_\ell - \OL{(\bu^{[p]})}_\ell\OL{(C^{[q]})}_\ell~.
\end{align}
\lb{eq:subscaleDecomp}
\end{subequations}
The term $\tau_\ell^{[p,q]}(\bu,C)$ represents the subscale contribution from scalar and velocity  scales within bands $[q]$ and $[p]$, respectively. Eyink \cite{Eyink05,Eyink06a} proved that the series~\eqref{eq:subscaleDecomp} converges absolutely and at a geometric rate under the condition that the scaling exponents of increments of $\delta\bu(r)\sim r^{\sigma^u}$ and $\delta C(r)\sim r^{\sigma^C}$ satisfy  $\sigma^u>0$ and $\sigma^C>0$ \footnote{More precisely, the condition is that $\sigma^u_p>0$ and $\sigma^C_p>0$ if increments scale as $\langle|\delta\bu(r)|^p\rangle^{1/p}\sim u_{\text{rms}}A_p (r/\ell_0)^{\sigma^u_p}$ and $\langle|\delta C(r)|^p\rangle^{1/p}\sim C_{\text{rms}}B_p (r/\ell_0)^{\sigma^C_p}$ for some dimensionless constants $A_p$ and $B_p$, where $\langle\dots\rangle$ is a space average. The condition has to hold for at  least $p=2$, with higher values of $p$ representing more stringent conditions. Note that a power-law scaling of increments is not required; the condition is still satisfied if increments decay faster than any power-law with $r\to 0$, for example exponentially.}. 

Exponents $\sigma^u$ and $\sigma^C$ reflect the smoothness of their respective fields (see Fig. 1 in \cite{Aluie17} and the associated discussion), with larger values of $\sigma$ implying smoother (or, heuristically, less turbulent) fields. For example, the Kolmogorov-Obukhov theory \cite{kolmogorov1941local,obukhov1949structure,corrsin1951spectrum} predicts $\sigma^u=\sigma^C=1/3$ for velocity and passive scalar increments over scales within the inertial–convective range of a high Reynolds number turbulent flow that is statistically homogeneous. In contrast, for low Reynolds number flows, even if they are chaotic, we can expect $\sigma^u\ge1$. Ignoring intermittency corrections, the scaling exponent $\sigma$ of increments is related to the power-law scaling of a spectrum, $E(k)$, of either $\bu$ or $C$, via $E(k)\sim k^{-2\sigma - 1}$. Therefore, the condition $\sigma>0$ corresponds to a wavenumber scaling of spectra that decays faster than $k^{-1}$ in wavenumber, which is a fairly weak condition and is expected to hold in most flows. 
The conditions are the same as those required for $\OL\tau_\ell(\bu,C)$ to be ultraviolet scale-local \cite{Eyink05,eyink2009localness,aluie2009localness}, which means that length-scales $\delta\ll\ell$ make a negligible contribution to $\OL\tau_\ell(\bu,C)$, 
\be
|\OL{\tau}_{\ell}(C'_\delta, \bu'_\delta)| \ll |\OL{\tau}_{\ell}(C, \bu)|.
\lb{eq:UVlocality}\ee

Under these conditions of convergence, the leading order term in eq.~\eqref{eq:subscaleDecomp} is the dominant contribution to the series, which justifies the approximation
\be
\OL\tau_\ell(\bu,C) \approx\tau_\ell^{[0,0]}(\bu,C) 
=\OL\tau_\ell(\OL{\bu}_\ell,\OL{C}_\ell). 
\lb{eq:LeadingTerm}\ee
This approximation corresponds to the well-known similarity (or Bardina) model \cite{Bardinaetal80,MeneveauKatz00}.
To readers who are perhaps less familiar with LES or coarse-graining, the term 
\be
\OL\tau_\ell(\OL{\bu}_\ell,\OL{C}_\ell)
= \OL{(\OL{\bu}_\ell\OL{C}_\ell)}_\ell
- \OL{(\OL{\bu}_\ell)}_\ell\OL{(\OL{C}_\ell)}_\ell
\lb{eq:Similarity}\ee
may seem insignificant, especially considering that if one were to replace operation $\OL{(\dots)}_\ell$ with a Reynolds (or ensemble) average, expression~\eqref{eq:Similarity} would be identically zero. However, operation $\OL{(\dots)}_\ell$ is a decomposition in length-scale, which is inherently different from Reynolds averaging. To aid in the conceptual understanding of expression~\eqref{eq:Similarity}, assume that $\OL{(\dots)}_\ell$ is a sharp-spectral projection in wavenumber space, $\OL{(\dots)}_\ell = (\dots)^{<K}$. Applying $(\dots)^{<K}$ removes all wavenumbers larger than $K=\ell^{-1}$ (\textit{i.e.} small length-scales) while keeping wavenumbers smaller than $K$ intact. Then 
\begin{subequations}
\begin{align}
\OL\tau_\ell(\OL{\bu}_\ell,\OL{C}_\ell)
&= (\bu^{<K}C^{<K})^{<K}
- (\bu^{<K})^{<K}(C^{<K})^{<K}
\lb{eq:Similarity_Sharp_1}\\    
&= (\bu^{<K}C^{<K})^{<K}
- \bu^{<K}C^{<K}
\lb{eq:Similarity_Sharp_2},  
\end{align}
\lb{eq:Similarity_Sharp}
\end{subequations}
using the sharp projection property, $((\dots)^{<K})^{<K} = (\dots)^{<K}$, to arrive at the last expression.
The term $\bu^{<K}C^{<K}$ in eq.\eqref{eq:Similarity_Sharp_2}, being a quadratic product, contains wavenumbers $<2K$. Of these, the contribution $(\bu^{<K}C^{<K})^{<K}$ from wavenumbers $<K$ is subtracted in eq.\eqref{eq:Similarity_Sharp_2}, such that $\OL\tau_\ell(\OL{\bu}_\ell,\OL{C}_\ell)$ represents wavenumbers within the band $[K,2K]$ or, equivalently, between length-scales $\ell$ and $\ell/2$. As discussed in previous work on scale-locality \cite{Eyink05,Aluie11}, this band of scales is expected to make the dominant contribution to the subscale flux $\OL\tau_\ell(\bu,C)$ if the spectrum decays faster than $k^{-1}$ in wavenumber over the range of scales $<\ell$. This is essentially the condition under which eq.~\eqref{eq:LeadingTerm} was derived. Physically, a spectral decay faster than $k^{-1}$ implies that dyadic wavenumber band $[k,2k]$ has more energy than band $[2k,4k]$ (e.g. \citep{frazier1991littlewood,AluieEyink09}), which justifies retaining the smallest wavenumbers via the leading order term in expansion~\eqref{eq:subscaleDecomp} \cite{aluie2009localness}.
A spectral decay faster than $k^{-1}$ is a fairly weak condition and is expected in most flows (turbulent or laminar), but may fail, for example, if $\ell$ approaches scales of the spectral peak or larger.
We emphasize that filtering with a sharp-spectral kernel in eq.~\eqref{eq:Similarity_Sharp} is merely for conceptual understanding. Unlike the Top-Hat or Gaussian, a sharp-spectral kernel is not positive semi-definite in physical space, which violates physical realizability conditions \cite{Vremanetal94}, rendering  subscale energy or coarse-grained mass density  negative in physical space \cite{Eyink05,Aluie13}.

\subsection{Relation to Increments and Gradients}
Using the usual definition of an increment,
\be\delta f(\bx;\br) = f (\mathbf{x}+\mathbf{r}) - f (\mathbf{x}),
\ee
the subscale flux \eqref{eq:subscaleflux} can be rewritten exactly in terms of $\delta C$ and $\delta \bu$ \cite{Constantinetal94,Eyink05,Aluie11}:
\begin{equation}
  \OL{\tau} (C, \bu)
    = \left\langle \delta {C} \, \delta
      {\bu} \right\rangle_{\ell}
    - \left\langle\delta C \right\rangle_{\ell}
      \left\langle \delta {\bu} \right\rangle_{\ell}.
\lb{eq:SubgridTauDeltas}\end{equation}
Equation \eqref{eq:SubgridTauDeltas} is exact and follows directly from eq.~\eqref{eq:subscaleflux} without requiring any assumption about scale-locality or the scaling exponents, where 
\be
\left\langle \delta {f} \right\rangle_{\ell}(\bx) \equiv \int  \mathrm{d}^{n}\br\, G_\ell(-\br) \delta {f}(\bx;\br)
\lb{eq:incrementsSubscales}\ee
 is a local average around $\bx$ over all separations $\br$ weighted by the kernel $G_\ell$. A spatially localized (or compact) kernel effectively limits the average in eq.~\eqref{eq:incrementsSubscales} to separations $|\br|\le\ell/2$. Note that $G_\ell(-\br) = G_\ell(\br)$ for even kernels, such as a Gaussian or a Top-Hat. Relation~\eqref{eq:SubgridTauDeltas} is a key step and reveals the connection between the subscale physics and increments (spatial variations) over distances smaller than $\ell$. This connection is underscored by noting that expression~\eqref{eq:incrementsSubscales} equals (minus) the high-pass field containing scales $<\ell$, 
 \be
 \int \mathrm{d}^{n}\br\, G_\ell(-\br) \, \delta {f}(\bx;\br) = -[f(\bx) - \OL{f}_\ell(\bx)] = -f'_\ell(\bx)~.
 \ee
We may also connect these considerations with the argument presented in the introduction, where increments reflect gradients \cite{Eyink06a}, including strain and vorticity of the subscale flow.
 
 Approximating the subscale flux by the leading order term in eq.\eqref{eq:LeadingTerm} allows us to use identity~\eqref{eq:SubgridTauDeltas} for the filtered quantities, 
 \begin{equation}
  \OL{\tau} (\OL{C}_{\ell}, \OL{\bu}_{\ell})
    = \left\langle \delta \OL{C}_{\ell} \, \delta
      \OL{\bu}_{\ell} \right\rangle_{\ell}
    - \left\langle\delta \OL{C}_{\ell} \right\rangle_{\ell}
      \left\langle \delta \OL{\bu}_{\ell} \right\rangle_{\ell}.
\lb{eq:SubgridTauDeltas_cg}\end{equation}
 Since a filtered field $\OL{f}_\ell(\bx)$ is smooth, we can 
Taylor expand its increments around $\bx$, 
\begin{equation}
  \delta \OL{f}_\ell(\bx;\br) = \OL{f}_\ell(\bx+\br)-\OL{f}_\ell(\bx)  \approx  \mathbf{r} \bdot \grad\OL{f}_\ell(\bx) + \dots,
\lb{eq:TaylorExpand}\end{equation}
where we neglect higher order terms.
Substituting the first term in the Taylor expansion of each of $\delta \OL{C}$ and $\delta \OL{\bu}$ into eq. \eqref{eq:SubgridTauDeltas_cg} gives in $n$-dimensions
\begin{subequations}
\begin{align}
  \OL{\tau}_\ell (\OL{C}, \OL{u}_i)
  &= \left[\partial_k\OL{C}_\ell\right]\left[\partial_m\OL{(u_i)}_\ell\right] \left[\left\langle r_k \,r_m  \right\rangle_{\ell} - \left\langle r_k \right\rangle_{\ell} \left\langle r_m  \right\rangle_{\ell}\right]\lb{eq:NLmodel_1}\\
  &= \left[\partial_k\OL{C}_\ell\right]\left[\partial_m\OL{(u_i)}_\ell\right] \left[\frac{1}{n}\delta_{km}\,\ell^2\int \mathrm{d}^n \mathbf{s} \, G (\mathbf{s}) \left | \mathbf{s}\right |^2  \right]\lb{eq:NLmodel_2}\\
  &= \frac{1}{n}  \, M_2\, \ell^2 \,\partial_k \OL{(u_i)}_\ell \,\partial_k \OL{C}_\ell~.\lb{eq:NLmodel_3}
\end{align}
\end{subequations}
In deriving the second line, we used  the symmetry of the kernel such that $\left\langle r_k \right\rangle_{\ell}=0$. In the final expression, the mother kernel's second moment, $M_2 = \int \mathrm{d}^n \mathbf{s} \, G (\mathbf{s}) \left | \mathbf{s}
\right |^2$,  depends solely on the shape of kernel $G$ and, in particular, is independent of scale $\ell$. For the Top-Hat in eq.~\eqref{eq:TopHat_mother},   $M_2=n/12$ in $n$-dimensions.

\subsection{Summary and Interpretation}
From the derivation above, we see that the final expression in eq.~\eqref{eq:NLmodel_3} represents the leading order contribution to $\OL{\tau}_\ell(C,\bu)$. Therefore, to the extent expression~\eqref{eq:NLmodel_3} faithfully approximates $\OL{\tau}_\ell(C,\bu)$, it also represents the subscale physics. How can expression~\eqref{eq:NLmodel_3}, which involves only coarse-grained (or resolved) quantities represent subscale interactions? It is analogous to inferring the value of a function $f(x_0+h)$ a distance $h$ away from $x_0$ based solely on its local properties at $x_0$ via a Taylor series expansion, the accuracy of which depends on smoothness properties of $f$. Here, the expansion is done in length-scale \cite{Eyink06a} rather than in space to arrive at eq.~\eqref{eq:NLmodel_3}. The analogue to smoothness is ultraviolet scale-locality, which is guaranteed under weak spectral scaling conditions \cite{Eyink05,eyink2009localness,aluie2009localness,Aluie11}.

Expression~\eqref{eq:NLmodel_3} corresponds to the nonlinear (or Clark) model \cite{clark1979evaluation}, which is a closure of the subscale flux using a product of gradients in LES modeling and has been shown to be an excellent (\emph{a priori}) approximation of the subscales in several previous studies (e.g. \cite{Liuetal94,porte2001priori,higgins2004heat}). Some studies have also offered viable ideas for utilizing it in \textit{a posteriori} LES \cite{Bardinaetal80,kosovic1997subgrid,bouchet2003parameterization}. Below, we provide additional \emph{a priori} support from simulations of 3D compressible turbulence and two-layer stacked shallow water simulations of geophysical fluid turbulence.

In the discussion above, we followed Eyink's multiscale gradient expansion \cite{Eyink06a} to derive expression~\eqref{eq:NLmodel_3}, which is more general than the standard derivations of the nonlinear model found in the LES literature (e.g. \cite{clark1979evaluation,Bardinaetal80,Liuetal94,leonard1997large,BorueOrszag98,Pope00,bouchet2003parameterization}). Since Eyink's multiscale gradient expansion is convergent, it provides a rigorous foundation to justify neglecting higher order terms. Another advantage of using Eyink's expansion is in bringing to the fore the underlying physics, including the assumptions under which the series~\eqref{eq:subscaleDecomp} converges and, therefore, the conditions under which the approximation may fail. Moreover, being a systematic expansion, it allows for future improvements by retaining higher order terms or by using a closed form of those terms following the recent work of Johnson \cite{johnson2020energy,johnson2021role}.

Approximation~\eqref{eq:NLmodel_3} provides us with physical insight into the mechanisms by which the subscales transport $\OL{C}_\ell$. Below, we show that subscale vorticity's contribution is an eddy-induced drift velocity. Prior to doing so, we discuss how well eq.~\eqref{eq:NLmodel_3} approximates $\OL\tau_\ell(\bu,C)$ from simulations.

\section{Empirical Support}\lb{sec:Empirical}
Approximating $\OL\tau_\ell(\bu,C)$ by expression~\eqref{eq:NLmodel_3} has significant support in the LES literature from \emph{a priori} tests of the nonlinear model. It has been shown to exhibit excellent agreement when $C$ is taken as the momentum (e.g. \cite{Liuetal94,BorueOrszag98}), as the temperature field in the atmospheric boundary layer \cite{porte2001priori,higgins2004heat}, and as a passive scalar \cite{kang2001passive,chumakov2008priori}. 

In this section, we present additional supporting evidence for active scalars, when $C$ represents (i) density in a 3D compressible turbulent flow and (ii) layer thickness in a two-layer stacked shallow water simulation of geophysical fluid turbulence. The correlation is generally better than $0.9$ at inertial scales smaller than the scale at which the spectrum peaks.

\subsection{Compressible Turbulence}\lb{sec:DiNuSUR_3D1024}
We carry out a direct numerical simulation (DNS) of forced compressible turbulence in a periodic box of non-dimensional size $2 \pi$ with grid resolution $1024^3$. From the simulation data, we test how well the expression in eq.~\eqref{eq:NLmodel_3} approximates $\OL\tau_\ell(\bu,C)$ with $C$ replaced by the density field, $\rho$, which is an active scalar.

\begin{figure}
    \centering
    \includegraphics[width = \textwidth]{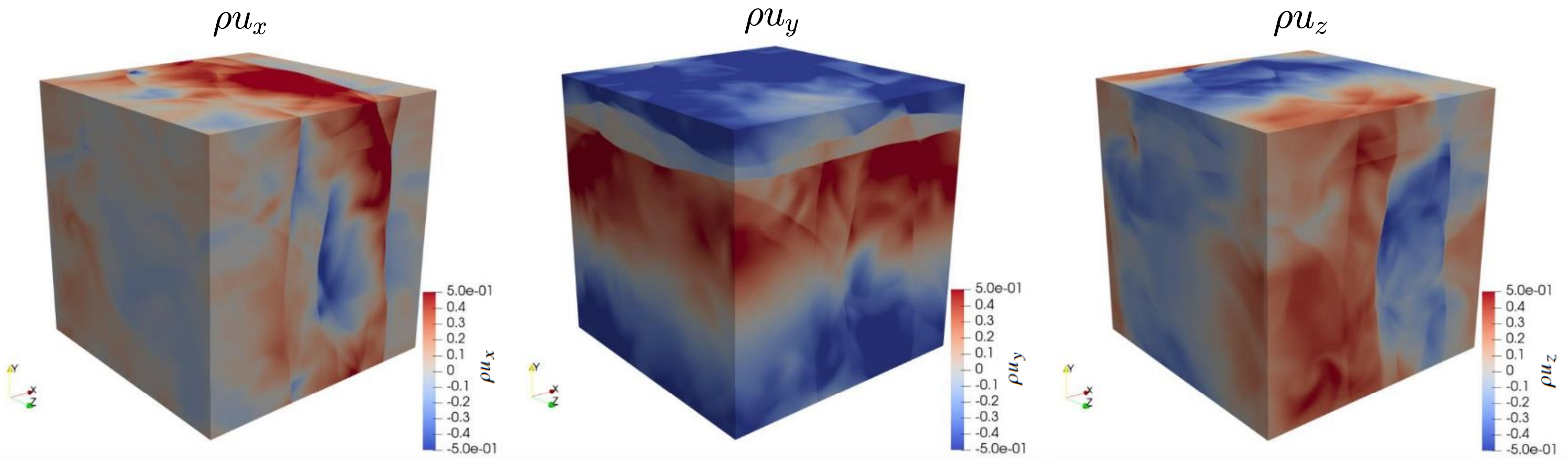}
    \caption{Left to right: visualization of x, y and z components, respectively, of momentum, $\rho \, \bu$, in physical space from the $1024^3$ DNS. These images are shown at an instant of time after the flow has reached steady state. Shocks can be seen as discontinuities. The spectrum of this flow is shown in Fig.~\ref{fig:Spectrum} in the Appendix.}
    \label{fig:PG}
\end{figure}
The DNS follows the configuration of \cite{lees2019baropycnal} and solves the fully compressible Navier-Stokes equations using our DiNuSUR code \cite{ZhaoAluie18,lees2019baropycnal}:
\begin{eqnarray} 
&\hspace{-0.4cm}\partial_t \rho& + \partial_j(\rho \, u_j) = 0, \lb{continuity} \\
&\hspace{-0.4cm}\partial_t (\rho \, u_i)& + \partial_j(\rho \, u_i \, u_j) 
= -\partial_i P +  \partial_j\sigma_{ij} + \rho \, F_i~,   \lb{momentum}\\
&\hspace{-0.4cm}\partial_t (\rho \, E)&+ \partial_j(\rho \, E \, u_j) 
= -\partial_j (P \, u_j) +\partial_j[2 \, \mu \, u_i \, (S_{ij} - \frac{1}{3} S_{kk} \, \delta_{ij})]  -\partial_j q_j  +\rho \, u_i \, F_i - \mathcal{RL}~. \lb{total-energy}\end{eqnarray}
Here, $E=|\bu|^2/2 + e$ is total energy per unit mass, where $e$ is specific internal energy,  $P$ is thermodynamic pressure, $\mu$ is dynamic viscosity, ${\bf q} = -\kappa \grad T$ is the heat flux with a thermal conductivity $\kappa$ and temperature $T$.  Both dynamic viscosity and thermal conductivity are spatially variable, where $\mu(\bx) = \mu_0 (T(\bx)/T_0)^{0.76}$. Thermal conductivity is set to satisfy a Prandtl number $Pr = c_p \, \mu/\kappa =0.7$, where $c_p=R\,\gamma/(\gamma-1)$ is the specific heat with specific gas constant $R$ and $\gamma=5/3$ for a monoatomic gas. We use the ideal gas equation of state, $P=\rho \, R \, T$. 
We stir the flow using an external acceleration field $F_i$, and $\mathcal{RL} = \rho \, u_i \, F_i$ represents radiation losses from internal energy (see \cite{JagannathanDonzis16,lees2019baropycnal}). 
$S_{ij} = (\partial_j u_i + \partial_i u_j)/2$ is the symmetric rate of strain tensor and $\sigma_{ij}$ is the deviatoric (traceless) viscous stress, 
\be \sigma_{ij}=2 \, \mu \, (S_{ij} - \frac{1}{3} S_{kk} \, \delta_{ij}).
\lb{eq:viscousstress}\ee 
We solve the above equations using the pseudo-spectral method with $2/3$rd dealiasing. We advance in time using the $4^{th}$-order Runge-Kutta scheme with a variable time step.

The acceleration, $\bF$, is similar to that in \cite{Federrathetal08}. 
In Fourier space, the acceleration is defined as
\begin{equation}
  \widehat{F}_i(\mathbf{k}) = \widehat{f}_j (\mathbf{k})
    \, P_{ij}^\zeta (\mathbf{k}),
\end{equation}
where $\mathbf{\widehat{f}}$ is constructed from 
independent Ornstein-Uhlenbeck stochastic processes \cite{EswaranPope88}. The
projection operator 
\be
P_{ij}^\zeta(\mathbf{k}) = \zeta \, \delta_{ij} +
(1 - 2 \, \zeta) \, \frac{k_i \,  k_j}{|\mathbf{k}|^2}
\ee
allows for controlling the ratio of
solenoidal ($\grad \bdot \mathbf{F} = 0$) to dilatational
($\grad \btimes \mathbf{F} = 0$) components of the forcing using the parameter
$\zeta$. When $\zeta = 0$, the forcing is purely dilatational and when $\zeta = 1$,
the forcing is purely solenoidal. For the simulation we use here, shown in Fig.~\ref{fig:PG}, we set $\zeta=0.01$ to simulate turbulence with high compressibility, such that dilatation, $\grad\bdot\bu \ne 0$, is significant. 

Our simulated flow has significant compressibility, which can be seen from Fig.~\ref{fig:PG} and  inferred from the following three time-averaged quantities. 
(1) The ratio $(\grad \bdot \bu)_{rms}/(\grad \btimes \bu)_{rms}= 5.61$ in our flow, which measures compressibility at small scales.
(2) We also Helmholtz decompose the velocity field into the dilatational, $\grad \btimes \bu^d = 0$, and solenoidal, $ \grad \bdot\bu^s = 0$, components, $\bu = \bu^d + \bu^s$. The ratio of dilatational kinetic energy, $K^d = \langle \rho \, u^d_i \, u^d_i/2 \rangle$, to solenoidal kinetic energy, $K^s = \langle \rho \, u^s_i \, u^s_i/2 \rangle$, yields a measure of compressibility at large scales, which in our flow is $K^d/K^s = 4.80$. 
(3) We also calculate the turbulence Mach number, $M_t=\langle|\bu|^2\rangle^{1/2}/c = 0.23$, where $c=2.18$ is the mean non-dimensional sound speed. Spectra of density and velocity are plotted in Fig.~\ref{fig:Spectrum} in the Appendix. 

\begin{figure}
    \centering
    \includegraphics[width = \textwidth]{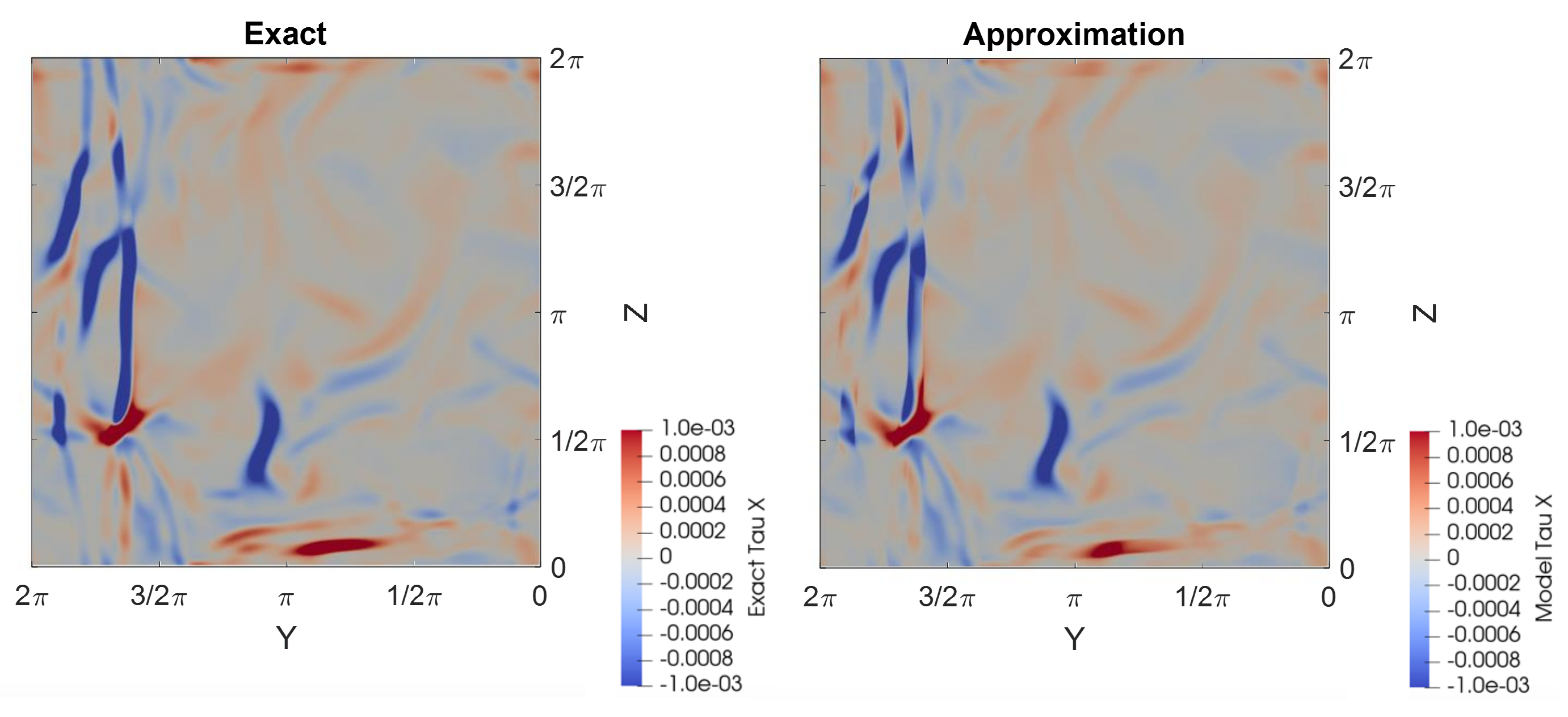}
    \caption{A 2D slice at $x = 0$ from a snapshot of the 3D compressible turbulence DNS, comparing the (left) exact $\OL{\tau}_\ell(u_x,\rho)$ at $\ell=0.19635$ with (right) its approximation from eq.~\eqref{eq:NLmodel_3}, $\tau_m = \frac{1}{3}M_2 \, \ell^2 \, \partial_k \OL\rho \, \partial_k \OL{u_x}$.}
    \label{fig:TauVizX}
\end{figure}

\begin{figure}
    \centering
    \includegraphics[width = \textwidth]{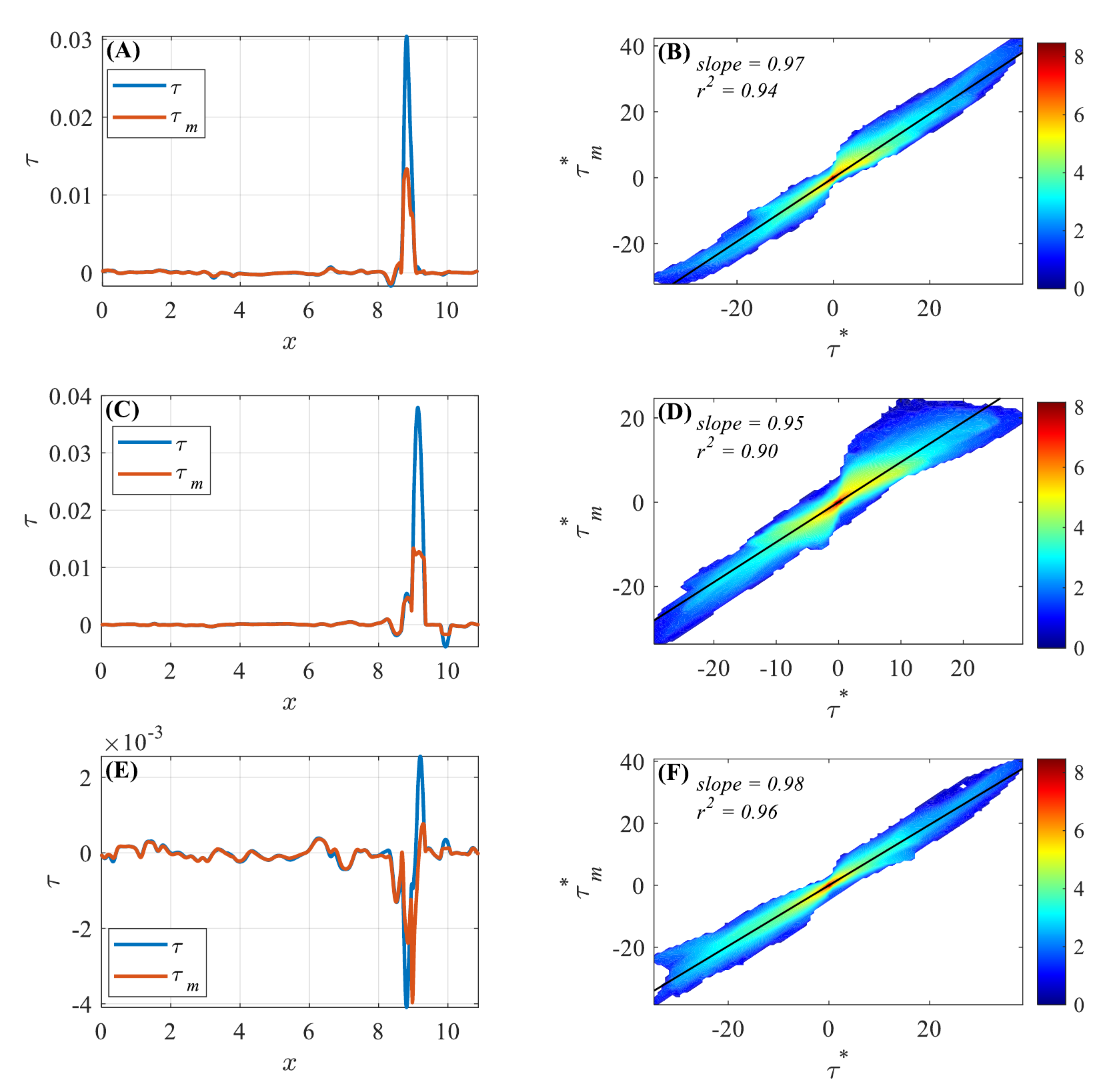}
    \caption{Comparison of $\OL\tau_\ell(\rho, \bu)$ with its approximation $\tau_m = \frac{1}{3}M_2 \, \ell^2 \, \partial_k \OL\rho \, \partial_k \OL\bu$ at $\ell = 0.19635$. (A) x-component of $\tau$ and $\tau_m$ along a diagonal transect through the domain from a single snapshot. (B) Joint PDF between $\tau^*$ and $\tau_m^*$, where superscript `$*$' indicates z-scores, which emphasizes trends (see eq.~\eqref{eq:Zscore} and associated discussion in the text). Color bar indicates the number of grid-points on a logarithmic scale, proportional to the probability.  Black line is the best linear fit, and the correlation-squared, $r^2$, quantifies the fraction of variability in $\tau$ that is captured by $\tau_m$. (C), (D) Same as (A), (B) but for the y-component. (E), (F) Same as (A), (B) but for the z-component.}
    \label{fig:CorrTauRhoU_k32}
\end{figure}

We use the Top-Hat kernel in eq.~\eqref{eq:TopHat} for the coarse-graining. Fig.~\ref{fig:TauVizX} shows a 2D slice from the 3D flow at the instant of time shown in Fig.~\ref{fig:PG}, comparing the exact $\OL{\tau}_\ell(u_x,\rho)$ at $\ell=0.19635$ (domain size is $2\pi$) with its approximation \eqref{eq:NLmodel_3}. The other two components are shown in the Appendix Fig.~\ref{fig:TauVizY}. All indicate an excellent agreement. Fig.~\ref{fig:CorrTauRhoU_k32} (left column) shows a more quantitative evaluation of all three components of the subscale transport approximation, where we can see that the approximation is remarkably accurate almost everywhere but underestimates $\OL{\tau}_\ell(u_x,\rho)$ at locations of strong shocks. This underestimation indicates that higher order terms in the expansion \eqref{eq:subscaleDecomp} are needed for better agreement at those locations since these rare but strong discontinuities have a significant contribution from scales $<\ell$ and are an extreme test case for the approximation. Indeed, at a location $x_0$ of a shock, the increment scales as $\delta u(x_0;r) = u(x_0+r) - u(x_0)  \sim r^\sigma$, with $\sigma\approx0$ for distances $r$ larger than the viscous shock width (see Fig.~1 in \cite{Aluie17} and the associated discussion). This value is at the margin of validity of the $\sigma>0$ condition \footnote{Formally, the condition $\sigma>0$ for Eyink's expansion to converge is not based on the scaling of an increment $\delta u(x;r)$ at a single point $x$. Rather, it is based on the scaling of structure functions, which are spatial averages of increments.} for Eyink's expansion \eqref{eq:subscaleDecomp} to converge.

Fig.~\ref{fig:CorrTauRhoU_k32} (right column) shows the joint probability density function (PDF) between $\OL{\tau}_\ell(\bu,\rho)$ and its approximation, with correlations exceeding $0.9$. In Figs.~\ref{fig:CorrTauRhoU_k16}-\ref{fig:CorrTauRhoU_k4} in the Appendix, we show similar comparisons at other length-scales. The correlation coefficient between two datasets, $A$ and $B$, is defined as
\begin{equation}
r(A,B) = \frac{\langle A\, B \rangle-\langle A \rangle\langle B \rangle}{[(\langle A^2 \rangle-\langle A \rangle^2)(\langle B^2 \rangle-\langle B \rangle^2)]^\frac{1}{2}}~,
\end{equation}
where $\langle\dots\rangle$ is a space-time average over the entire domain and time after which the flow has become statistically stationary. The joint PDF in Fig.~\ref{fig:CorrTauRhoU_k32} uses the so-called `z-scores' of $\tau_m$ and $\OL\tau_\ell$, denoted by $\tau^*_m$ and $\tau^*$ (all other figures show the actual $\OL\tau_\ell$ and $\tau_m$ values). The z-score of A is defined as
\be
A^* = \frac{A-\langle A\rangle}{\langle (A-\langle A\rangle)^2 \rangle^{1/2}},
\lb{eq:Zscore}\ee
which helps focus on the spatio-temporal variation patterns after subtracting the mean and normalizing by the standard deviation. For completeness, we plot the joint PDFs of the actual $\tau_m$ and $\OL\tau_\ell$ in Fig.~\ref{fig:CorrTauRhoU_k32-nostar} in the Appendix. Using z-scores in the joint PDF only affects the slope of the linear fit between $\tau_m$ and $\OL\tau_\ell$.
The correlation coefficient is the same whether using the actual values or their z-scores, 
\be r(A^*,B^*) = r(A,B).
\lb{eq:CorrZscores}\ee
Therefore, z-scores show how well the approximation $\tau_m$ is able to capture the exact subscale physics $\OL\tau_\ell$ up to a proportionality constant. 

\begin{figure}
    \centering
    \includegraphics[width = \textwidth]{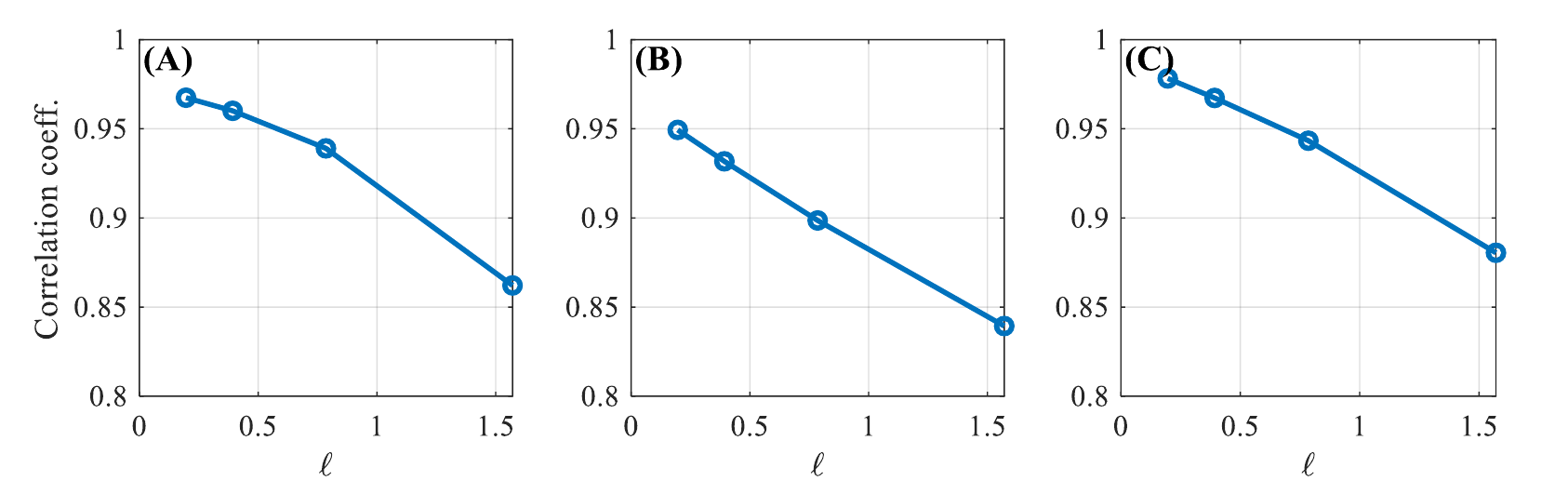}
    \caption{Correlation coefficient $r$ between $\OL\tau_\ell(\rho, \bu)$ and its approximation $\tau_m = \frac{1}{3}M_2 \, \ell^2 \, \partial_k \OL\rho \, \partial_k \OL\bu$ as a function of scale $\ell$. (A) x-component. (B) y-component. (C) z-component.}
    \label{fig:CorrTauRhoU_Ell}
\end{figure}

Fig.~\ref{fig:CorrTauRhoU_Ell} plots the correlation coefficient as a function of $\ell$, which has values exceeding $0.90$ for scales $\ell$ smaller than that of the peak and decreasing to $\approx 0.85$ for scales $\ell=\pi$ (half the domain size). These strong correlations justify our usage of expression~\eqref{eq:NLmodel_3} as a proxy for $\OL{\tau}_\ell(\bu,\rho)$ to gain insight into its underlying mechanics.

\subsection{Geophysical fluid turbulence}
We also ran simulations of the two-layer stacked shallow water equations~\eqref{momEqn},\eqref{thickness} using the ocean general circulation model MOM6 \cite{adcroft2019gfdl,griffies2020primer}. We simulated an ocean current \cite{hallberg2013using} that is baroclinically unstable  \cite{phillips1954energy,vallisatmospheric}, resulting in mesoscale eddy generation and geostrophic turbulence, which we visualize in Fig.~\ref{PhillipsFlowViz}. 

Details of the simulation setup can be found in \cite{hallberg2013using}. The equations solved in MOM6 are
\begin{equation}
    \frac{\partial \boldsymbol{u}_m}{\partial t} + (f + \omega)\,\hat{\bf{z}} \times \boldsymbol{u}_m = -\grad (M_m + |\boldsymbol{u}_m|^2/2 ) - \grad \bdot \boldsymbol{T} -\delta_{m2} \, C_D \, |\boldsymbol{u}_2| \, \boldsymbol{u}_2~,
    \label{momEqn}
\end{equation}
\begin{equation}
    \frac{\partial h_m}{\partial t} + \grad \bdot(h_m \bu_m) = (3 - 2m)\left[\gamma\left(\OL{\eta_{3/2}}^x - \eta_{3/2, ref}\right) - \grad\bdot\left(K_h\grad\eta_{3/2}\right)\right]~.
    \label{thickness}
\end{equation}
The subscript $m = 1,2$ labels the top and bottom layers, respectively, and repeated indices are not summed in the thickness equation \eqref{thickness}. $\eta_{1/2}$ is the vertical position of the top layer interface (surface height), $\eta_{3/2}$ is the height of the interface between the two layers,  $\boldsymbol{u}_m$ is the horizontal velocity, $\omega$ is vertical component of relative vorticity,  $f = f_0 + \beta y$ is the Coriolis parameter for Earth's rotation (i.e., the planetary vorticity), where $f_0$ is the value of $f$ at southernmost boundary ($y = 0$) and $\beta$ is the linear rate of change of $f$ along meridional direction. $\boldsymbol{T}$ is the horizontal stress tensor parameterized by Smagorinsky biharmonic viscosity \cite{griffies2000biharmonic}, $C_D$ is the dimensionless bottom drag coefficient and $M_1 = g \, \eta_{1/2}$ and $M_2 = M_1 +  \left(g\frac{\Delta\rho}{\rho_0}\right)\eta_{3/2}$ are the pressure of the layers 1 and 2 normalized by $\rho_0$, the density of the top layer. $\Delta\rho$ is the density difference between the two layers, $h_m$ is the layer thickness, $K_h$ is the interface height diffusivity coefficient, and $\gamma$ is a rate that is proportional to the mass flux between the layers \cite{hallberg2013using}. $K_h$ allows for the dissipation of available potential energy \cite{gent1990isopycnal} and $\gamma$ forces the interface height between the two layers to reference zonal mean state $\eta_{3/2, ref}$. Here, $\OL{(\cdot)}^x$ represents a zonal average (i.e., in the east-west direction). The flow is able to attain a statistically steady state in the absence of direct wind forcing in eq.~\eqref{momEqn} by nudging the layer interface $\eta_{3/2}$ to the reference profile $\eta_{3/2, ref}$ in eq.~\eqref{thickness}, which drives the flow by providing a source of potential energy \cite{hallberg2013using}, mimicking the sloping of isopycnals by Ekman pumping in the real ocean \cite{vallisatmospheric}.

\begin{figure}
    \centering
    \includegraphics{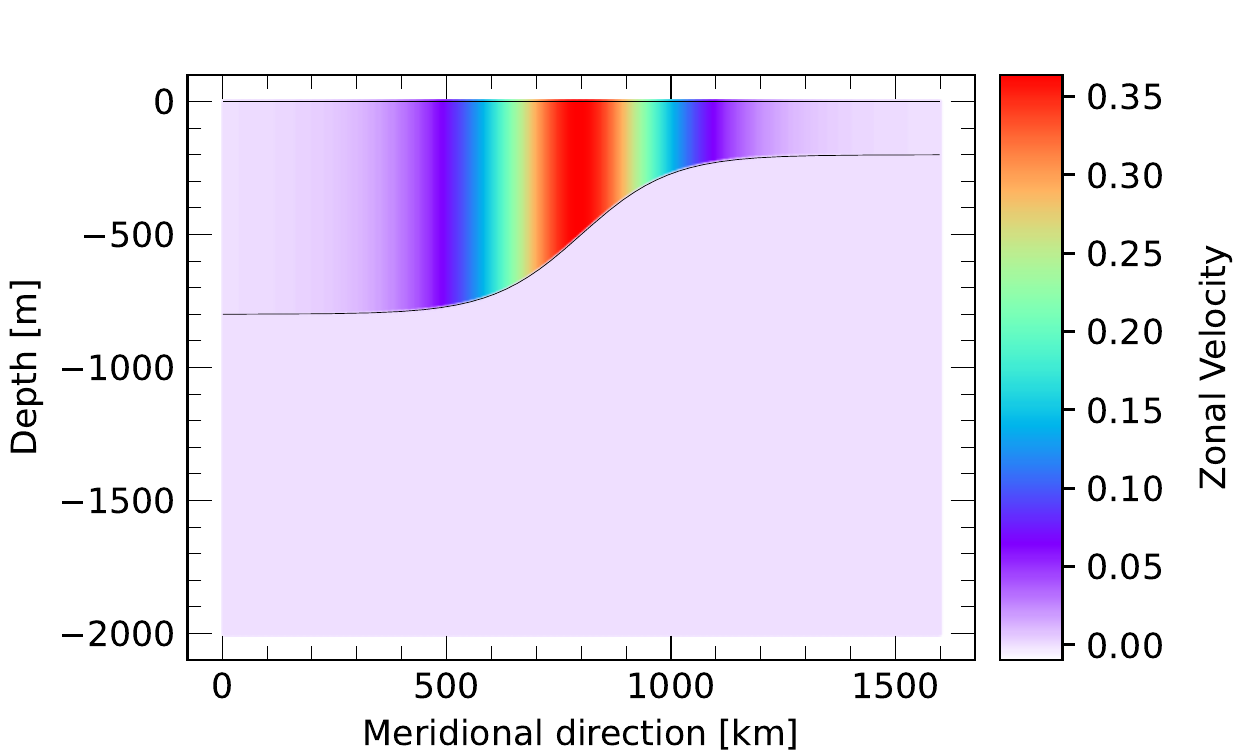}
    \caption{Initial condition for two-layer stacked shallow water simulation following \cite{hallberg2013using}. The top layer has zonal (east-west) velocity with a Gaussian profile peaking at $\approx0.35~$m/s (positive to the east) while the bottom layer is initially quiescent. Bottom topography is flat. The surface has initial height $\eta_{1/2}=0~$m. The interface height $\eta_{3/2}$ between the two layers is initialized to the reference $\eta_{3/2,ref}$  shown to satisfy geostrophic balance. $\eta_{3/2}$ is allowed to evolve freely while being relaxed back to $\eta_{3/2,ref}$ at the rate $\gamma$.}
    \label{fig:PhillipsIC}
\end{figure}

The flat bottom domain is $1200~$km $\times$ $1600~$km in horizontal extent and is discretized on a Cartesian grid with a grid-cell size of 5~km. The total depth is $2~$km and the bottom surface is flat. It is zonally periodic and subject to free-slip walls at the northern and southern boundaries. The initial conditions have a zonal jet $200~$km wide with a Gaussian profile peaking at $\approx0.35~$m/s for top layer and bottom layer is quiescent, similar to initial conditions of \cite{hallberg2013using}. The interface between the two layers is such that it satisfies geostrophic balance and is as shown in figure \ref{fig:PhillipsIC}.  The Coriolis parameter at the southernmost boundary is $f_0=6.49 \times 10^{-5} s^{-1}$ and $\beta = \partial_{y}f = 2 \times 10^{-11} m^{-1}s^{-1}$. The value of $\gamma = 1.1574 \times 10^{-6} ~ s^{-1}$ and $K_h =8000~m^2s^{-1} $. 
\begin{figure}
    \centering
    \includegraphics[width=\textwidth]{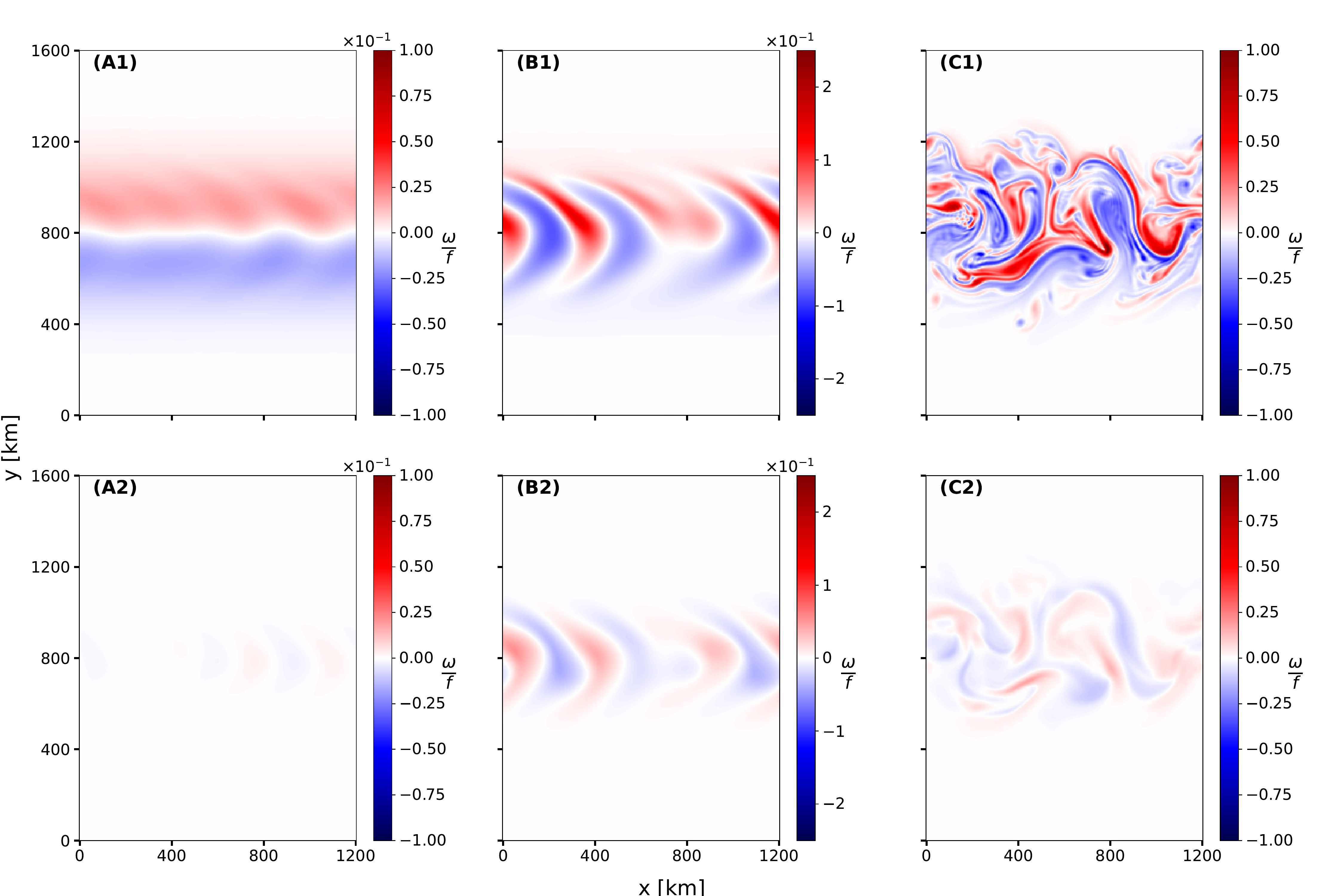}
    \caption{Visualization of the relative vorticity normalized by the Coriolis paramater $f$ at the domain's mid-latitude from our simulation of  an eastward flowing ocean current that is baroclinically unstable, resulting in eddy generation and turbulence.
    The simulation solves the two-layer shallow water equations, with the top and bottom panels showing the respective top and bottom layers. Time increases from left to right: panels A1 and A2 are from day 60, B1 and B2 from day 93, C1 and C3 from day 180. Our analysis uses model data after day 169, when the flow has become statistically steady.}
    \label{PhillipsFlowViz}
\end{figure}
The simulation is run for 180 days, sufficiently long for the geostrophic turbulence to develop as a result of baroclinic instabilities. For our results, we analyze 100 snapshots taken every 2.5 hours starting from day 169. Spectra of $\bu_1$ and $h_1$ are shown in Appendix Fig.~\ref{fig:KEandh2spectra}. For our purpose, we only analyze the fields from the top layer since the flow in the bottom layer is much weaker. Hereafter, we drop the layer subcript since all results pertain to the top layer.

We compare $\OL\tau_\ell(h,\bu)$
with its approximation $\frac{1}{2} \ell^2 \, M_2 \, \partial_k\OL{h} \,\partial_k\OL{\bu}$ in eq.~\eqref{eq:NLmodel_3}. We use a graded Top-Hat kernel following our previous work analyzing oceanic flow \cite{aluie2018mapping,Raietal2021}, 
\begin{equation}
    G_\ell(r) = 0.5 - 0.5 \tanh((r - \ell/2)/10),
    \label{weight}
\end{equation}
where $\ell$ denotes the kernel width. This kernel is essentially the same as the Top-Hat in eq.\eqref{eq:TopHat}, but with smoothed edges to avoid discretization noise from the nonuniform grids \cite{aluie2018mapping} commonly used in general circulation models. Since the model domain is flat, $r$ is a simple Euclidean distance as in eq.~\eqref{eq:TopHat}, without complications due to Earth's surface curvature \cite{aluie2019convolutions}. 

\begin{figure}
    \centering
    \includegraphics[width = \textwidth]{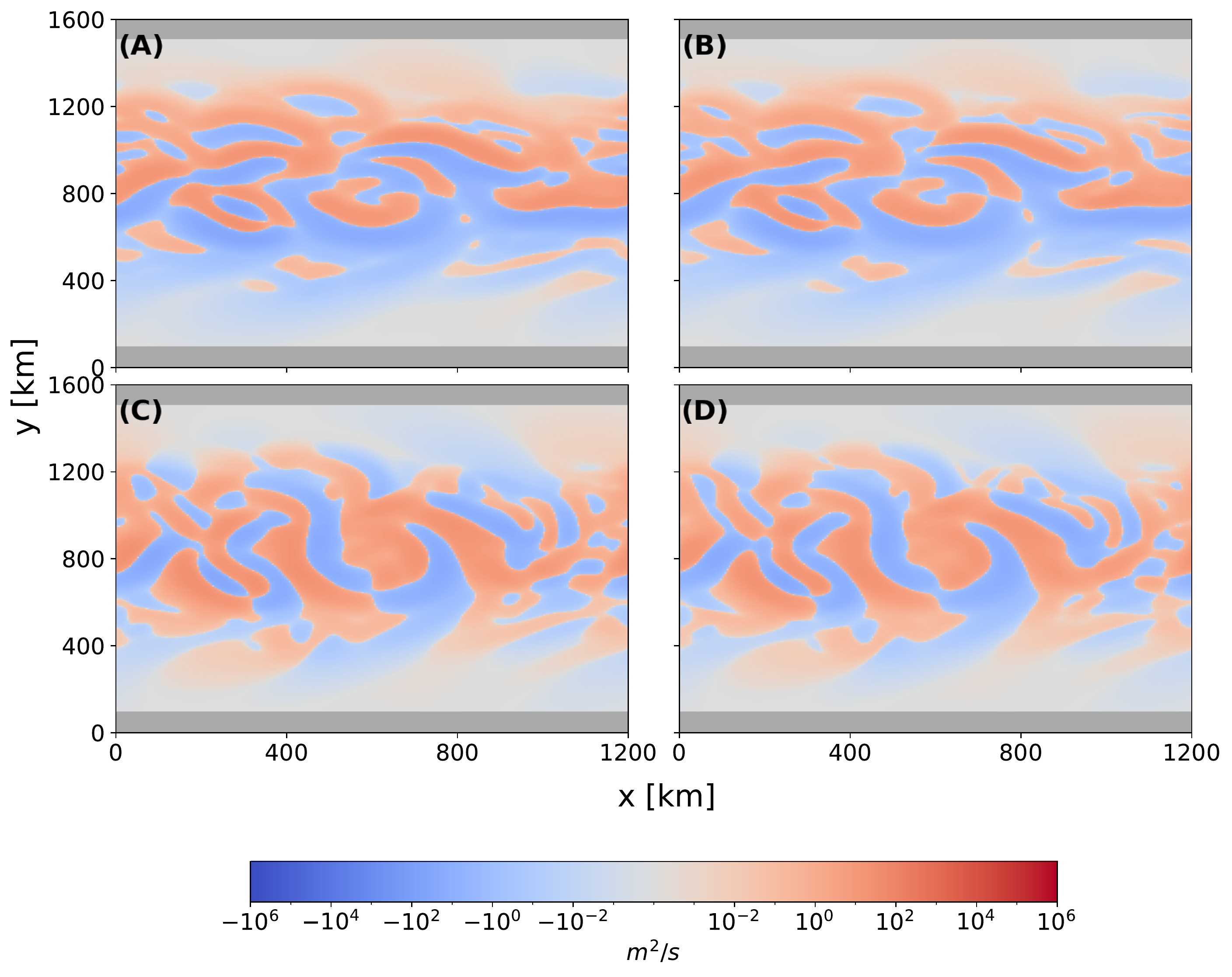}
    \caption{Comparing the (left) exact $\OL{\tau}_\ell(\bu,h)$ at $\ell=100~$km with (right) its approximation from eq.~\eqref{eq:NLmodel_3}, $\tau_m = \frac{1}{2}M_2 \, \ell^2 \, \partial_k \OL{h} \, \partial_k \OL{\bu}$, from a single snapshot. Top panels are for the zonal (eastward) component. Bottom panels are for the meridional (northward) component. Grey bands at northern and southern boundaries are excluded from our analysis to avoid details of the coarse-graining next to boundaries.}
    \label{fig:pmesh100km}
\end{figure}
\begin{figure}
    \centering
    \includegraphics[width = \textwidth]{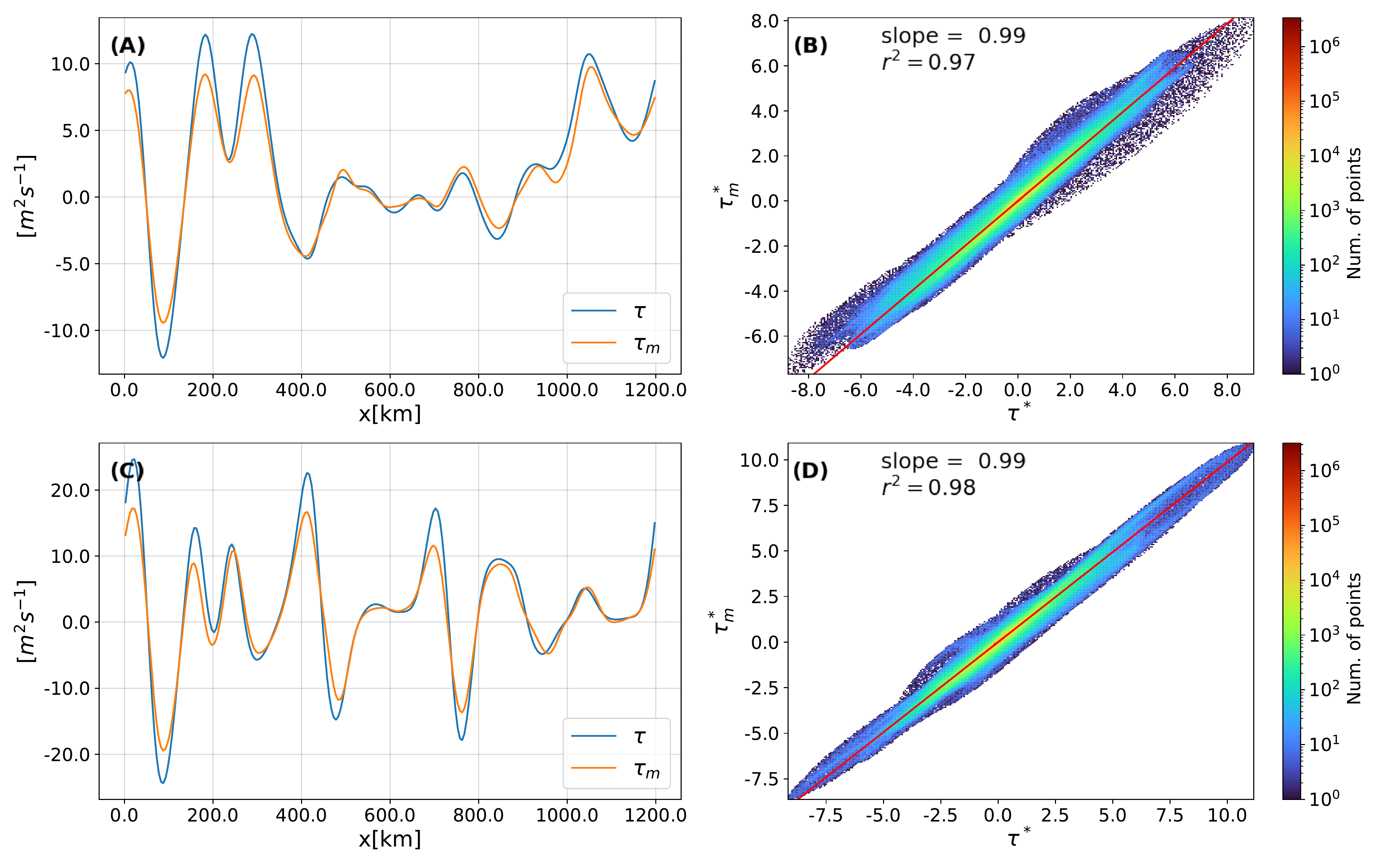}
    \caption{ Comparison of $\OL\tau_\ell(h, \bu)$ with its approximation $\tau_m = \frac{1}{2}M_2 \, \ell^2 \, \partial_k \OL{h} \, \partial_k \OL\bu$ at $\ell = 100~$km. (A) Zonal (eastward) component of $\tau$ and $\tau_m$ along the transect $y = 802.5~$km from a snapshot on day 171 of the simulation. (B) Joint PDF for all space-time points between $\tau^*$ and $\tau_m^*$, where superscript `$*$' indicates z-scores, which emphasizes trends (see eq.~\eqref{eq:Zscore} and associated discussion in the text). Color bar indicates the number of grid-points on a logarithmic scale, proportional to the probability.  Red line is the best linear fit, and the correlation-squared, $r^2$, quantifies the fraction of variability in $\tau$ that is captured by $\tau_m$.}
    \label{fig:corr100km}
\end{figure}

Fig.~\ref{fig:pmesh100km} compares the $\OL{\tau}_\ell(\bu,h)$ at $\ell=100~$km with its approximation \eqref{eq:NLmodel_3}. It indicates an excellent agreement. Fig.~\ref{fig:corr100km} (left column) shows a more quantitative evaluation of both components of the subscale transport approximation, where we can see that the approximation is remarkably accurate almost everywhere.
Fig.~\ref{fig:corr100km} (right column) shows the joint PDFs between $\OL{\tau}_\ell(\bu,\rho)$ and its approximation for both components (eastward and northward directions). They show correlations exceeding $0.95$. In Figs.~\ref{fig:corr50km}-\ref{fig:corr400km} in the Appendix, we show similar comparisons at other length-scales of 50~km, 200~km, and 400~km. 
The joint PDF in Fig.~\ref{fig:corr100km} uses the z-scores of $\OL\tau_\ell$ and $\tau_m$, denoted by $\tau^*$ and $\tau^*_m$ and  (all other figures show the actual $\OL\tau_\ell$ and $\tau_m$ values). As mentioned in eq.~\eqref{eq:CorrZscores}, the correlation coefficient between $\tau^*$ and $\tau^*_m$ is the same as that between $\OL\tau_\ell$ and $\tau_m$. We use z-scores to focus on how well the approximation $\tau_m$ is able to capture the exact subscale physics $\OL\tau_\ell$ up to a proportionality constant. For completeness, we plot the joint PDFs of the actual $\tau_m$ and $\OL\tau_\ell$ in Fig.~\ref{fig:pdensityPlotWithoutNormalization} in the Appendix. 

\begin{figure}
    \centering
    \includegraphics[width = \textwidth]{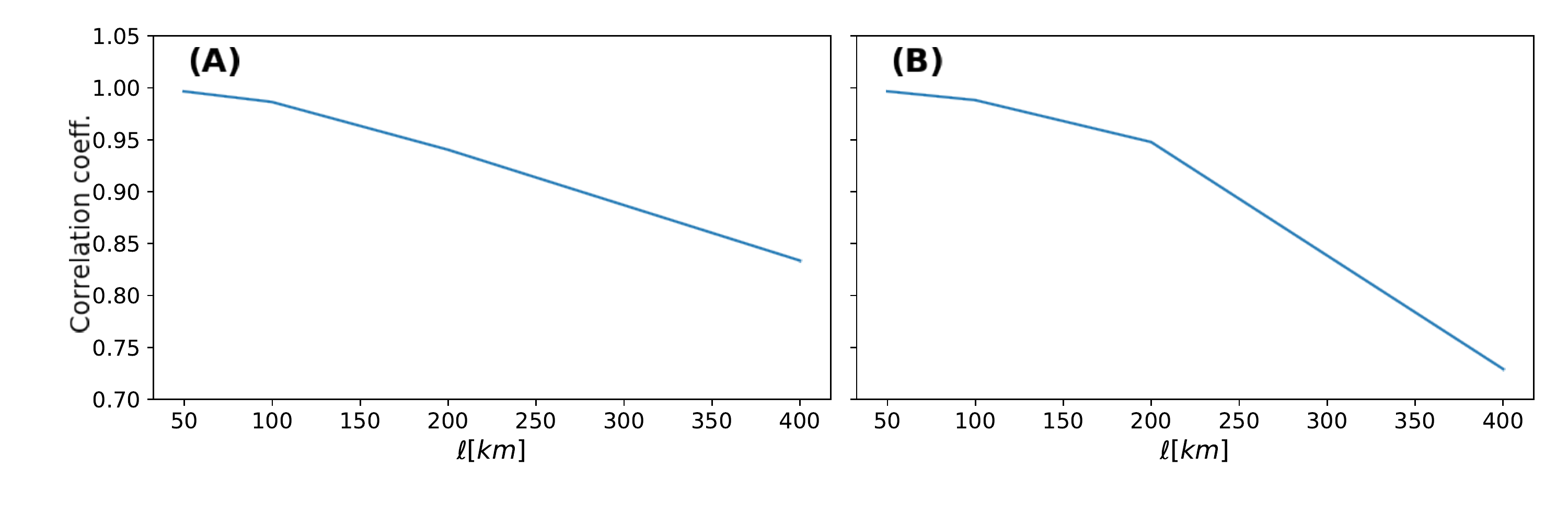}
    \caption{Correlation coefficient $r$ between $\OL\tau_\ell(h, \bu)$ and its approximation $\tau_m = \frac{1}{2}M_2 \, \ell^2 \, \partial_k \OL{h} \, \partial_k \OL\bu$ as a function of scale $\ell$. (A) Zonal (eastward) direction. (B) Meridional (northward) direction.}
    \label{fig:corrAsFuncOfEll}
\end{figure}

Noting that the spectral peak (Fig.~\ref{fig:KEandh2spectra}) is at $\approx300~$km, Fig.~\ref{fig:corrAsFuncOfEll} plots the correlation coefficients between the exact $\OL{\tau}_\ell(\bu,h)$ and its approximation~\eqref{eq:NLmodel_3} as a function of $\ell$, which has values exceeding $0.90$ for scales $\ell$ smaller than the that of the spectral peak and decreasing to $\approx0.75$ for scales $\ell=400~$km. These high correlation values justify our usage of expression~\eqref{eq:NLmodel_3} as a proxy for $\OL{\tau}_\ell(\bu,h)$ to gain insight into its underlying mechanics at least for scales $\ell$ smaller than that of the spectral peak.

\section{Transport by Subscale Vorticity and Strain}\lb{sec:MainResult}
Insight from eq.~\eqref{eq:NLmodel_3} into how the subscale physics contributes to large-scale transport can be gleaned by using a fundamental result from \cite{Moffatt1983}, \cite{MiddletonLoder1989},
and \cite{griffies1998gent}, who showed in different physical systems the  importance of scalar transport by an anti-symmetric skew diffusion tensor, or equivalently an eddy-induced advective flux. We elaborate on this idea in this section, which contains the main result of the paper in eq.~\eqref{eq:EffectVel}. 

Replacing $\OL\tau_\ell(\bu,C)$ in eq.~\eqref{eq:CoarseAdvect} by expression~\eqref{eq:NLmodel_3}, we have
\be \partial_t \OL{C}_\ell + \grad\bdot (\OL{\bu}_\ell \,\OL{C}_\ell) =   \grad\bdot\left(- A\, \ell^2 \, \partial_j \OL{\bu}_\ell \, \partial_j \OL{C}_\ell \right),
\lb{eq:CoarseAdvect_NLmodel}\ee
with the constant $A=M_2/n$. Therefore, transport of $\OL{C}_\ell$ by the subscale flux,
\be R(C) =  \grad\bdot [  \underbrace{\left(- A~ \ell^2\partial_j \OL{\bu}_\ell \right)}_{K_{ij}} \, \partial_j \OL{C}_\ell ],
\ee
resembles that of generalized transport   tensor 
\begin{equation}
  K_{ij} 
  = -A \, \ell^2 \, \partial_j \overline{u}_{i}. 
\end{equation}
Note that $K_{ij}$ is a function of the length scale $\ell$ through both the $\ell^{2}$ factor as well as the coarse-grained velocity $\OL{\bu}_\ell$. For notational brevity we drop the $\ell$ when writing individual components of $\OL{\bu}_\ell$.

Physical properties of the subscale transport tensor $K_{ij}$ can be illuminated by separating it into a symmetric part $K^{\mbox{\tiny sym}}_{ij}$ representing subscale strain, and anti-symmetric part (also called skew-symmetric) $K^{\mbox{\tiny skew}}_{ij}$ representing subscale vorticity. The tensor is rewritten as
\be K_{ij} = - A \, \ell^2 \, \partial_j \OL{u}_{i} = - A~ \ell^2 \, [ \underbrace{\frac{1}{2}\left(\partial_j \OL{u}_i + \partial_i \OL{u}_j\right)}_{\OL{S}_{ij}} + \underbrace{\frac{1}{2}\left(\partial_j \OL{u}_i - \partial_i \OL{u}_j\right)}_{\OL\Omega_{ij}}
 ] = K^{\mbox{\tiny sym}}_{ij} + K^{\mbox{\tiny skew}}_{ij}
\ee
with vorticity and strain underlying kinematically distinct flow processes \citep[e.g.,][]{griffies1998gent}. Note that the symmetric part satisfies $K^{\mbox{\tiny sym}}_{ij} = K^{\mbox{\tiny sym}}_{ji}$ while the anti-symmetric part satisfies $K^{\mbox{\tiny skew}}_{ij} = -K^{\mbox{\tiny skew}}_{ji} $.

Transport due to the subscale strain, 
\begin{eqnarray} 
R^{\mbox{\tiny sym}}(C) = \partial_i [  K^{\mbox{\tiny sym}}_{ij} \, \partial_j \OL{C}_\ell ], 
\end{eqnarray}
can be thought of in the local eigenframe reference of $K^{\mbox{\tiny sym}}_{ij}$ as anisotropic diffusion/anti-diffusion of $\OL{C}_\ell$ in directions that are not necessarily parallel to $\grad \OL{C}_\ell$. Such transport\footnote{The tensor $K^{\mbox{\tiny sym}}$ would have to be positive semi-definite to qualify as a diffusion tensor in the strict sense.} was analyzed in detail in \cite{higgins2004heat}. Positive eigenvalues are associated with a stretching flow direction (diffusion) while negative eigenvalues are associated with a contracting flow direction (anti-diffusion). If the underlying flow is non-solenoidal, $\grad\bdot\bu\ne0$, such as in our simulations presented here, then $K^{\mbox{\tiny sym}}_{ij}$ can also account for isotropic diffusion or coalescence due to the subscale physics. 

To understand the role of subscale vorticity, represented by
the anti-symmetric (or skew-symmetric) tensor $K^{\mbox{\tiny skew}}_{ij}$, we follow the analysis of \cite{Moffatt1983,MiddletonLoder1989,griffies1998gent} and rewrite
\begin{eqnarray} 
R^{\mbox{\tiny skew}}(C) = \partial_i [  K^{\mbox{\tiny skew}}_{ij} \, \partial_j \OL{C}_\ell ] 
= [\partial_i   K^{\mbox{\tiny skew}}_{ij}]  \partial_j \OL{C}_\ell =\partial_j [\underbrace{\partial_i   K^{\mbox{\tiny skew}}_{ij}}_{-(\bv_*)_j}   \OL{C}_\ell]
\end{eqnarray}
where the second equality follows from the chain rule and noting that the product of $K^{\mbox{\tiny skew}}_{ij}$ (antisymmetric tensor) with $\partial_i\partial_j \OL{C}_\ell$ (symmetric tensor) vanishes. Similarly, the third equality follows from $\partial_i \partial_j   K^{\mbox{\tiny skew}}_{ij} =0$. The interpretation of $\bv_*$ is an eddy-induced velocity arising from the stirring induced by subscale vortical motions. It is straightforward to show that 
\be \bv_*=-\grad\bdot{\bf K}^{\mbox{\tiny skew}} = \frac{1}{2} \, A \, \ell^2 \,\grad\btimes(\grad\btimes\OL{\bu}_\ell),
\lb{eq:EffectVel}\ee
which is the main result of this paper. An example of how it can arise in swirling flow is sketched in Fig.~\ref{fig:EffectVort}. Note that this eddy-induced velocity is solenoidal, 
\begin{equation}
   \grad\bdot\bv_* = 0, 
\end{equation} 
due to the antisymmetry of $ K^{\mbox{\tiny skew}}_{ij}$. 

\subsection{Summary and Interpretation}
The above result is a manifestation of the equivalence between transport by (i) flux due to an anti-symmetric diffusion tensor and (ii) an advective flux
\citep{Moffatt1983,MiddletonLoder1989,griffies1998gent}.
It suggests that coarse-grained simulations such as LES may need to solve 
\begin{eqnarray} \partial_t \OL{C}_\ell + \grad\bdot \left[(\OL{\bu}_\ell+\bv_*)\OL{C}_\ell \right] = -\grad\bdot\left( {\bf J}(C) \right) 
\lb{eq:CoarseAdvect_NLmodel_2}
\end{eqnarray}
to represent the unresolved (subgrid) vorticity physics self-consistently. 

${\bf J}(C)$ in eq.~\eqref{eq:CoarseAdvect_NLmodel_2} can represent traditional subgrid models such as turbulent (Smagorinsky) diffusion \cite{smagorinsky1963general}, ${\bf J}(C) = -\alpha_{turb}\grad \OL{C}_\ell$. The evaluation of $\bv_* \sim \grad\btimes(\grad\btimes\OL{\bu}_\ell)$ presents no additional cost since the resolved velocity $\OL\bu_\ell$ is known. Figs.~\ref{fig:v_star_OL_u_PDF}-\ref{fig:pdfVStarVsOL_v} in the Appendix show the joint PDFs of $\bv_*$ and $\OL{\bu}_\ell$ at different length-scales $\ell$, which demonstrate that $\bv_*$ and the coarse velocity $\OL{\bu}_\ell$ are often of a comparable magnitude.

Our interpretation of $\bv_*$ as an eddy-induced velocity representing subscale vortical motions can be justified by appealing to a relation by Johnson\footnote{Johnson's relation~\eqref{eq:JohnsonIdentity} is exact only when filtering with a Gaussian kernel.} \cite{johnson2020energy,johnson2021role}, 
\be
\OL\tau_\ell(u_i,C) = 2\int_0^{\ell} d\delta\, \delta\, \OL{\left(\partial_k \OL{(u_i)}_{\delta} \, \partial_k\OL{C}_{\delta}\right)}_{\sqrt{\ell^2-\delta^2}}~~.
\lb{eq:JohnsonIdentity}
\ee
Eq.~\eqref{eq:JohnsonIdentity} highlights that subscale flux $\OL\tau_\ell(\bu,C)$ is due to the cumulative contribution of subscale velocity and scalar gradients at all scales $<\ell$. Since subscale velocity gradients  can be decomposed into strain and vorticity, and since expression~\eqref{eq:NLmodel_3} is the dominant contribution, we are justified in interpreting $\bv_*$ as an eddy-induced velocity representing subscale vortical motions.

\section{Conclusion}
We have shown that unlike subscale strain, which acts as an anisotropic diffusion/anti-diffusion tensor, subscale vorticity's contribution at leading order is solely a conservative advection of coarse-grained scalars by an eddy-induced velocity $\bv_*$ proportional to the curl of vorticity. Our analysis relied on the leading order term in Eyink's multiscale gradient expansion, which coincides with the nonlinear model and is known to be an excellent \emph{a priori} representation of the subscale physics. Here, we provided additional evidence for this excellent agreement from a three-dimensional compressible turbulence simulation and from a two-layer shallow water model of geophysical turbulence. Since the convergence of Eyink's expansion and, therefore, the dominance of the leading order term relies on ultraviolet scale-locality, our results and conclusions may not hold at length-scales larger than those of the spectral peak.

While the focus of this paper was on the transport of scalars, a similar analysis may also apply to the transport of momentum. Indeed, this is straightforward for incompressible flows where approximation~\eqref{eq:NLmodel_3} has been shown to work very well for momentum transport\cite{clark1979evaluation,Liuetal94,BorueOrszag98}. Johnson's recent work on the energy cascade \cite{johnson2021role} seems to indicate that approximation~\eqref{eq:NLmodel_3} may be deficient in its magnitude (by a factor $\approx\frac{1}{2}$) compared to the true momentum flux. Yet, since the two exhibit excellent spatio-temporal correlation as was shown in \cite{Liuetal94,BorueOrszag98}, a simple rescaling of expression~\eqref{eq:NLmodel_3} by a space-time independent coefficient (possibly $\approx2$) would suffice to match the true momentum flux. Once approximation~\eqref{eq:NLmodel_3} (possibly $\times2$) is accepted as representative of the subscale flux, the same argument we presented here follows through for momentum after replacing $C$ with $\bu$ in eq.~\eqref{eq:NLmodel_3}. In flows with significant density variations such as compressible turbulence, the subscale flux of momentum has a somewhat different functional form than in eq.~\eqref{eq:subscaleflux} due to the density-weighted filtering required to satisfy the inviscid criterion\cite{Aluie13,livescu2020turbulence}. Yet, we conjecture that a result similar to what we derived here is still possible for subscale momentum flux in variable-density flows, which was shown to be scale-local in \cite{Aluie11}.

\section*{Acknowledgement}
We thank Brandon Reichl at GFDL and the reviewers for valuable feedback and suggestions. This research was funded by US DOE grant DE-SC0020229. Partial support from US NSF grants PHY-2020249, OCE-2123496, US NASA grant 80NSSC18K0772, and US NNSA grant DE-NA0003856 is acknowledged. HA was also supported by US DOE grants DE-SC0014318, DE-SC0019329, NSF grant PHY-2206380, and US NNSA grant DE-NA0003914. JKS was also supported by US DOE grant DE-SC0019329, US NNSA grant DE-NA0003914, and US NSF grant CBET-2143702. Computing time was provided by NERSC under Contract No. DE-AC02-05CH11231 and NASA's HEC Program through NCCS at Goddard Space Flight Center. MOM6 is publicly available and can be accessed at \url{https://github.com/mom-ocean/MOM6}. The coarse-graining analysis was performed using codes similar to FlowSieve, which is publicly available at \url{https://github.com/husseinaluie/FlowSieve}. Other data used here may be obtained from the corresponding author upon reasonable request.

\clearpage
\newpage

\bibliographystyle{unsrt}
 \bibliography{SubscaleFlux}

\begin{thebibliography}{10}

\bibitem{Eyink06a}
Gregory~L Eyink.
\newblock {Multi-scale gradient expansion of the turbulent stress tensor}.
\newblock {\em Journal of Fluid Mechanics}, 549(-1):159, February 2006.

\bibitem{kundu2015fluid}
Pijush~K Kundu, Ira~M Cohen, and David~R Dowling.
\newblock {\em Fluid mechanics}.
\newblock Academic press, 2015.

\bibitem{Pope00}
S.~B. Pope.
\newblock {\em Turbulent flows}.
\newblock Cambridge University Press, New York, 2000.

\bibitem{MeneveauKatz00}
C.~{Meneveau} and J.~{Katz}.
\newblock {Scale-Invariance and Turbulence Models for Large-Eddy Simulation}.
\newblock {\em Ann. Rev. Fluid Mech.}, 32:1--32, 2000.

\bibitem{Eyink05}
G.~L. {Eyink}.
\newblock {Locality of turbulent cascades}.
\newblock {\em Physica D}, 207:91--116, 2005.

\bibitem{Strichartz03}
R~S Strichartz.
\newblock {\em {A guide to distribution theory and Fourier transforms}}.
\newblock World Scientific Publishing Company, 2003.

\bibitem{Evans10}
Lawrence~C Evans.
\newblock {\em {Partial Differential Equations}}.
\newblock Amer Mathematical Society, April 2010.

\bibitem{Leonard74}
A.~{Leonard}.
\newblock {Energy Cascade in Large-Eddy Simulations of Turbulent Fluid Flows}.
\newblock {\em Adv. Geophys.}, 18:A237, 1974.

\bibitem{Germano92}
M~Germano.
\newblock Turbulence: the filtering approach.
\newblock {\em Journal of Fluid Mechanics}, 238:325--336, 1992.

\bibitem{daubechies1992ten}
Ingrid Daubechies.
\newblock {\em Ten lectures on wavelets}.
\newblock SIAM, 1992.

\bibitem{Aluie17}
Hussein Aluie.
\newblock {Coarse-grained incompressible magnetohydrodynamics: analyzing the
  turbulent cascades}.
\newblock {\em New Journal of Physics}, January 2017.

\bibitem{eyink2018review}
Gregory~L Eyink.
\newblock Review of the onsager" ideal turbulence" theory.
\newblock {\em arXiv preprint arXiv:1803.02223}, 2018.

\bibitem{grinstein2007implicit}
Fernando~F Grinstein, Len~G Margolin, and William~J Rider.
\newblock {\em Implicit large eddy simulation}, volume~10.
\newblock Cambridge university press Cambridge, 2007.

\bibitem{higgins2004heat}
Chad~W Higgins, Marc~B Parlange, and Charles Meneveau.
\newblock The heat flux and the temperature gradient in the lower atmosphere.
\newblock {\em Geophysical research letters}, 31(22), 2004.

\bibitem{Aluie13}
H.~Aluie.
\newblock {Scale decomposition in compressible turbulence}.
\newblock {\em Physica D: Nonlinear Phenomena}, 247(1):54--65, March 2013.

\bibitem{ZhaoAluie18}
Dongxiao Zhao and Hussein Aluie.
\newblock {Inviscid criterion for decomposing scales }.
\newblock {\em Physical Review Fluids}, 3:054603, May 2018.

\bibitem{germano1986proposal}
Massimo Germano.
\newblock A proposal for a redefinition of the turbulent stresses in the
  filtered navier--stokes equations.
\newblock {\em The Physics of fluids}, 29(7):2323--2324, 1986.

\bibitem{kolmogorov1941local}
Andrey~Nikolaevich Kolmogorov.
\newblock The local structure of turbulence in incompressible viscous fluid for
  very large reynolds numbers.
\newblock {\em Dokl. Akad. Nauk SSSR}, 30(4):299--303, 1941.

\bibitem{obukhov1949structure}
Alexander~Mikhailovich Obukhov.
\newblock The structure of the temperature field in a turbulent flow.
\newblock {\em Izv. Akad. Nauk. SSSR, Ser. Geogr. and Geophys.}, 13:58, 1949.

\bibitem{corrsin1951spectrum}
Stanley Corrsin.
\newblock On the spectrum of isotropic temperature fluctuations in an isotropic
  turbulence.
\newblock {\em Journal of Applied Physics}, 22(4):469--473, 1951.

\bibitem{eyink2009localness}
Gregory~L Eyink and Hussein Aluie.
\newblock Localness of energy cascade in hydrodynamic turbulence. i. smooth
  coarse graining.
\newblock {\em Physics of Fluids}, 21(11):115107, 2009.

\bibitem{aluie2009localness}
Hussein Aluie and Gregory~L Eyink.
\newblock Localness of energy cascade in hydrodynamic turbulence. ii. sharp
  spectral filter.
\newblock {\em Physics of Fluids}, 21(11):115108, 2009.

\bibitem{Bardinaetal80}
J~Bardina, J~H Ferziger, and W~C Reynolds.
\newblock {Improved subgrid-scale models for large-eddy simulation}.
\newblock {\em American Institute of Aeronautics and Astronautics}, July 1980.

\bibitem{Aluie11}
H.~{Aluie}.
\newblock {Compressible Turbulence: The Cascade and its Locality}.
\newblock {\em Phys. Rev. Lett.}, 106(17):174502, April 2011.

\bibitem{frazier1991littlewood}
Michael Frazier, Michael~W Frazier, Bj{\"o}rn Jawerth, and Guido Weiss.
\newblock {\em Littlewood-Paley theory and the study of function spaces}.
\newblock Number~79 in Conference Board of the Mathematical Sciences regional
  conference series in mathematics. American Mathematical Society, Providence,
  Rhode Island, 1991.

\bibitem{AluieEyink09}
H.~{Aluie} and G.~{Eyink}.
\newblock {Localness of energy cascade in hydrodynamic turbulence. II. Sharp
  spectral filter}.
\newblock {\em Phys. Fluids}, 21(11):115108, November 2009.

\bibitem{Vremanetal94}
B.~{Vreman}, B.~{Geurts}, and H.~{Kuerten}.
\newblock {Realizability conditions for the turbulent stress tensor in
  large-eddy simulation}.
\newblock {\em J. Fluid Mech.}, 278:351--362, 1994.

\bibitem{Constantinetal94}
P.~{Constantin}, E.~{Weinan}, and E.~S. {Titi}.
\newblock {Onsager's conjecture on the energy conservation for solutions of
  Euler's equation}.
\newblock {\em Communications in Mathematical Physics}, 165:207--209, October
  1994.

\bibitem{clark1979evaluation}
Robert~A Clark, Joel~H Ferziger, and William~Craig Reynolds.
\newblock Evaluation of subgrid-scale models using an accurately simulated
  turbulent flow.
\newblock {\em Journal of fluid mechanics}, 91(1):1--16, 1979.

\bibitem{Liuetal94}
S~Liu, C~Meneveau, and J~Katz.
\newblock {On the properties of similarity subgrid-scale models as deduced from
  measurements in a turbulent jet}.
\newblock {\em Journal of Fluid Mechanics}, 275:83--119, 1994.

\bibitem{porte2001priori}
Fernando Port{\'e}-Agel, Marc~B Parlange, Charles Meneveau, and William~E
  Eichinger.
\newblock A priori field study of the subgrid-scale heat fluxes and dissipation
  in the atmospheric surface layer.
\newblock {\em Journal of the atmospheric sciences}, 58(18):2673--2698, 2001.

\bibitem{kosovic1997subgrid}
Branko Kosovi{\'c}.
\newblock Subgrid-scale modelling for the large-eddy simulation of
  high-reynolds-number boundary layers.
\newblock {\em Journal of Fluid Mechanics}, 336:151--182, 1997.

\bibitem{bouchet2003parameterization}
Freddy Bouchet.
\newblock Parameterization of two-dimensional turbulence using an anisotropic
  maximum entropy production principle.
\newblock {\em arXiv preprint cond-mat/0305205}, 2003.

\bibitem{leonard1997large}
A~Leonard.
\newblock Large-eddy simulation of chaotic convection and beyond.
\newblock In {\em 35th Aerospace Sciences Meeting and Exhibit}, page 204, 1997.

\bibitem{BorueOrszag98}
V~Borue and S~A Orszag.
\newblock {Local energy flux and subgrid-scale statistics in three-dimensional
  turbulence}.
\newblock {\em Journal of Fluid Mechanics}, 366(1), 1998.

\bibitem{johnson2020energy}
Perry~L Johnson.
\newblock Energy transfer from large to small scales in turbulence by
  multiscale nonlinear strain and vorticity interactions.
\newblock {\em Physical review letters}, 124(10):104501, 2020.

\bibitem{johnson2021role}
Perry~L Johnson.
\newblock On the role of vorticity stretching and strain self-amplification in
  the turbulence energy cascade.
\newblock {\em Journal of Fluid Mechanics}, 922, 2021.

\bibitem{kang2001passive}
Hyung~Suk Kang and Charles Meneveau.
\newblock Passive scalar anisotropy in a heated turbulent wake: new
  observations and implications for large-eddy simulations.
\newblock {\em Journal of Fluid Mechanics}, 442:161--170, 2001.

\bibitem{chumakov2008priori}
Sergei~G Chumakov.
\newblock A priori study of subgrid-scale flux of a passive scalar in isotropic
  homogeneous turbulence.
\newblock {\em Physical Review E}, 78(3):036313, 2008.

\bibitem{lees2019baropycnal}
Aarne Lees and Hussein Aluie.
\newblock Baropycnal work: a mechanism for energy transfer across scales.
\newblock {\em Fluids}, 4(2):92, 2019.

\bibitem{JagannathanDonzis16}
Shriram Jagannathan and Diego~A Donzis.
\newblock {Reynolds and Mach number scaling in solenoidally-forced compressible
  turbulence using high-resolution direct numerical simulations}.
\newblock {\em Journal of Fluid Mechanics}, 789:669--707, February 2016.

\bibitem{Federrathetal08}
Christoph Federrath, Ralf~S. Klessen, and Wolfram Schmidt.
\newblock The density probability distribution in compressible isothermal
  turbulence: Solenoidal versus compressive forcing.
\newblock {\em The Astrophysical Journal Letters}, 688(2):L79, 2008.

\bibitem{EswaranPope88}
V~Eswaran and SB~Pope.
\newblock An examination of forcing in direct numerical simulations of
  turbulence.
\newblock {\em Computers \& Fluids}, 16(3):257--278, 1988.

\bibitem{adcroft2019gfdl}
Alistair Adcroft, Whit Anderson, V~Balaji, Chris Blanton, Mitchell Bushuk,
  Carolina~O Dufour, John~P Dunne, Stephen~M Griffies, Robert Hallberg,
  Matthew~J Harrison, et~al.
\newblock The gfdl global ocean and sea ice model om4. 0: Model description and
  simulation features.
\newblock {\em Journal of Advances in Modeling Earth Systems},
  11(10):3167--3211, 2019.

\bibitem{griffies2020primer}
Stephen~M Griffies, Alistair Adcroft, and Robert~W Hallberg.
\newblock A primer on the vertical lagrangian-remap method in ocean models
  based on finite volume generalized vertical coordinates.
\newblock {\em Journal of Advances in Modeling Earth Systems},
  12(10):e2019MS001954, 2020.

\bibitem{hallberg2013using}
Robert Hallberg.
\newblock Using a resolution function to regulate parameterizations of oceanic
  mesoscale eddy effects.
\newblock {\em Ocean Modelling}, 72:92--103, 2013.

\bibitem{phillips1954energy}
Norman~A Phillips.
\newblock Energy transformations and meridional circulations associated with
  simple baroclinic waves in a two-level, quasi-geostrophic model.
\newblock {\em Tellus}, 6(3):274--286, 1954.

\bibitem{vallisatmospheric}
GK~Vallis.
\newblock Atmospheric and oceanic fluid dynamics: fundamentals and large-scale
  circulation, 2006.

\bibitem{griffies2000biharmonic}
Stephen~M Griffies and Robert~W Hallberg.
\newblock Biharmonic friction with a smagorinsky-like viscosity for use in
  large-scale eddy-permitting ocean models.
\newblock {\em Monthly Weather Review}, 128(8):2935--2946, 2000.

\bibitem{gent1990isopycnal}
Peter~R Gent and James~C Mcwilliams.
\newblock Isopycnal mixing in ocean circulation models.
\newblock {\em Journal of Physical Oceanography}, 20(1):150--155, 1990.

\bibitem{aluie2018mapping}
Hussein Aluie, Matthew Hecht, and Geoffrey~K Vallis.
\newblock Mapping the energy cascade in the north atlantic ocean: The
  coarse-graining approach.
\newblock {\em Journal of Physical Oceanography}, 48(2):225--244, 2018.

\bibitem{Raietal2021}
Shikhar Rai, Matthew Hecht, Matthew Maltrud, and Hussein Aluie.
\newblock Scale of oceanic eddy killing by wind from global satellite
  observations.
\newblock {\em Science Advances}, 7(28):eabf4920, 2021.

\bibitem{aluie2019convolutions}
Hussein Aluie.
\newblock Convolutions on the sphere: commutation with differential operators.
\newblock {\em GEM-International Journal on Geomathematics}, 10(1):9, 2019.

\bibitem{Moffatt1983}
H.K. Moffatt.
\newblock Transport effects associated with turbulence with particular
  attention to the influence of helicity.
\newblock {\em Reports on Progress in Physics}, 46:621--664, 1983.

\bibitem{MiddletonLoder1989}
J.~F. Middleton and J.~W. Loder.
\newblock Skew fluxes in polarized wave fields.
\newblock {\em Journal of Physical Oceanography}, 19:68--76, 1989.

\bibitem{griffies1998gent}
Stephen~M Griffies.
\newblock The gent--mcwilliams skew flux.
\newblock {\em Journal of Physical Oceanography}, 28(5):831--841, 1998.

\bibitem{smagorinsky1963general}
Joseph Smagorinsky.
\newblock General circulation experiments with the primitive equations: I. the
  basic experiment.
\newblock {\em Monthly weather review}, 91(3):99--164, 1963.

\bibitem{livescu2020turbulence}
Daniel Livescu.
\newblock Turbulence with large thermal and compositional density variations.
\newblock {\em Annual Review of Fluid Mechanics}, 52:309--341, 2020.

\bibitem{SadekAluie18}
Mahmoud Sadek and Hussein Aluie.
\newblock {Extracting the spectrum of a flow by spatial filtering}.
\newblock {\em Physical Review Fluids}, 3(12):124610, December 2018.

\bibitem{buzzicotti2021}
Michele Buzzicotti, Benjamin~A Storer, Stephen~M Griffies, and Hussein Aluie.
\newblock A coarse-grained decomposition of surface geostrophic kinetic energy
  in the global ocean.
\newblock {\em Earth and Space Science Open Archive}, page~58, 2021.

\end{thebibliography}
 
\clearpage
\newpage
 
\section*{Appendix}

\subsection{Additional supporting material from the compressible turbulence simulation}
\begin{figure}[h]
    \centering
    \includegraphics[width = \textwidth]{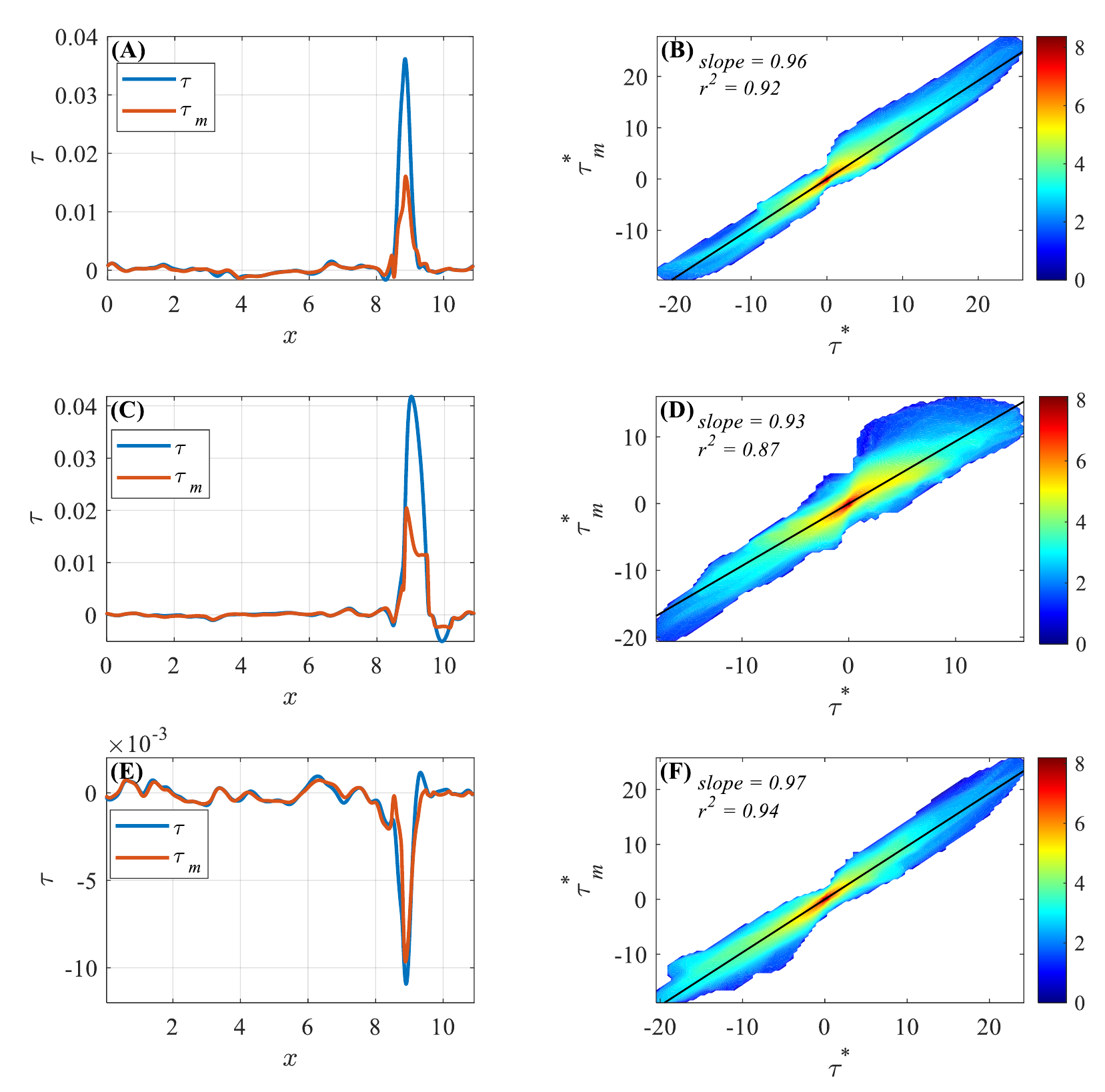}
    \caption{ Same as in Fig. \ref{fig:CorrTauRhoU_k32} for $\ell = 0.3927$.}
    \label{fig:CorrTauRhoU_k16}
\end{figure}

\begin{figure}
    \centering
    \includegraphics[width = \textwidth]{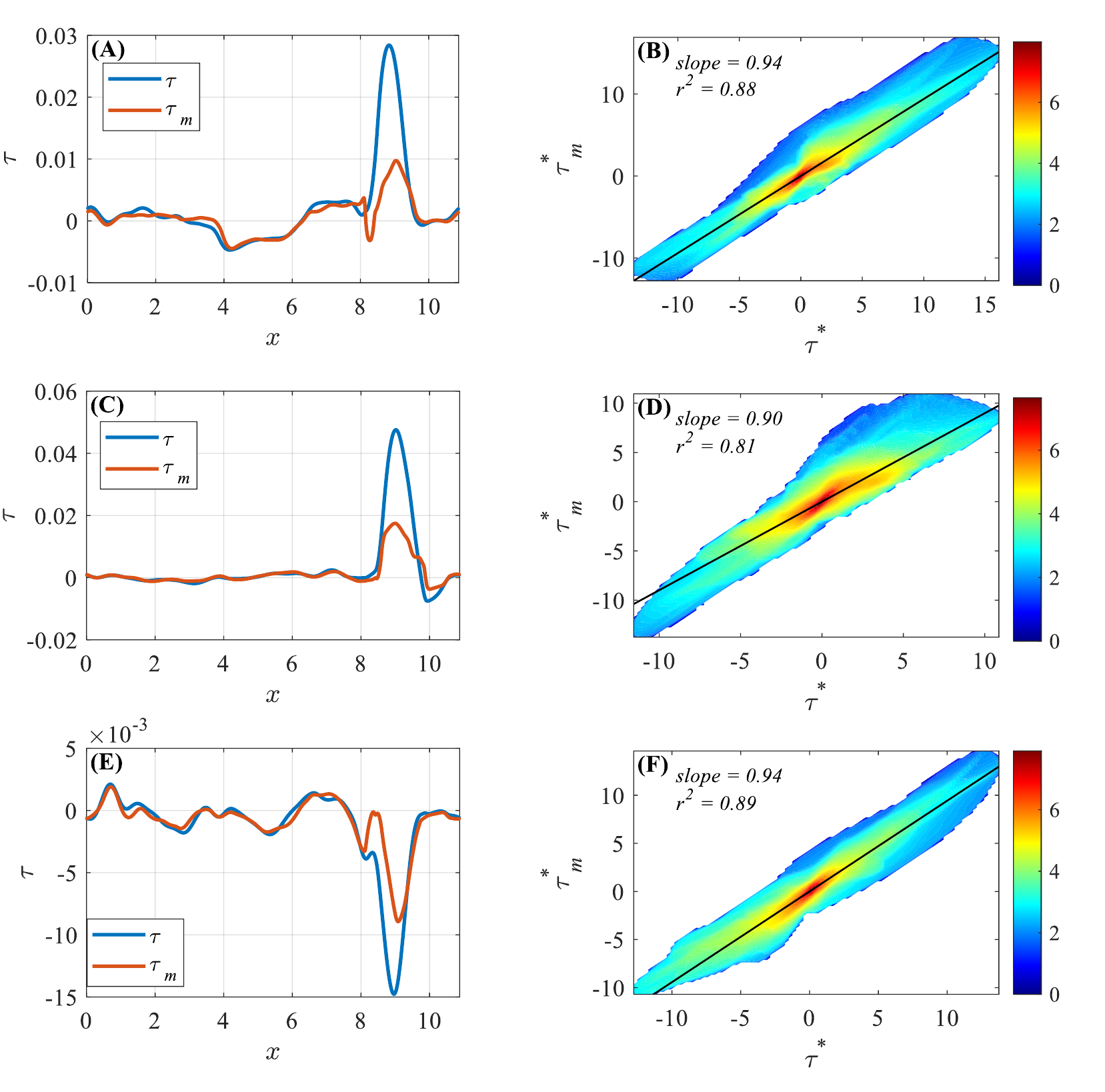}
    \caption{ Same as Fig. \ref{fig:CorrTauRhoU_k32} for $\ell = 0.7854$.}
    \label{fig:CorrTauRhoU_k8}
\end{figure}

\begin{figure}
    \centering
    \includegraphics[width = \textwidth]{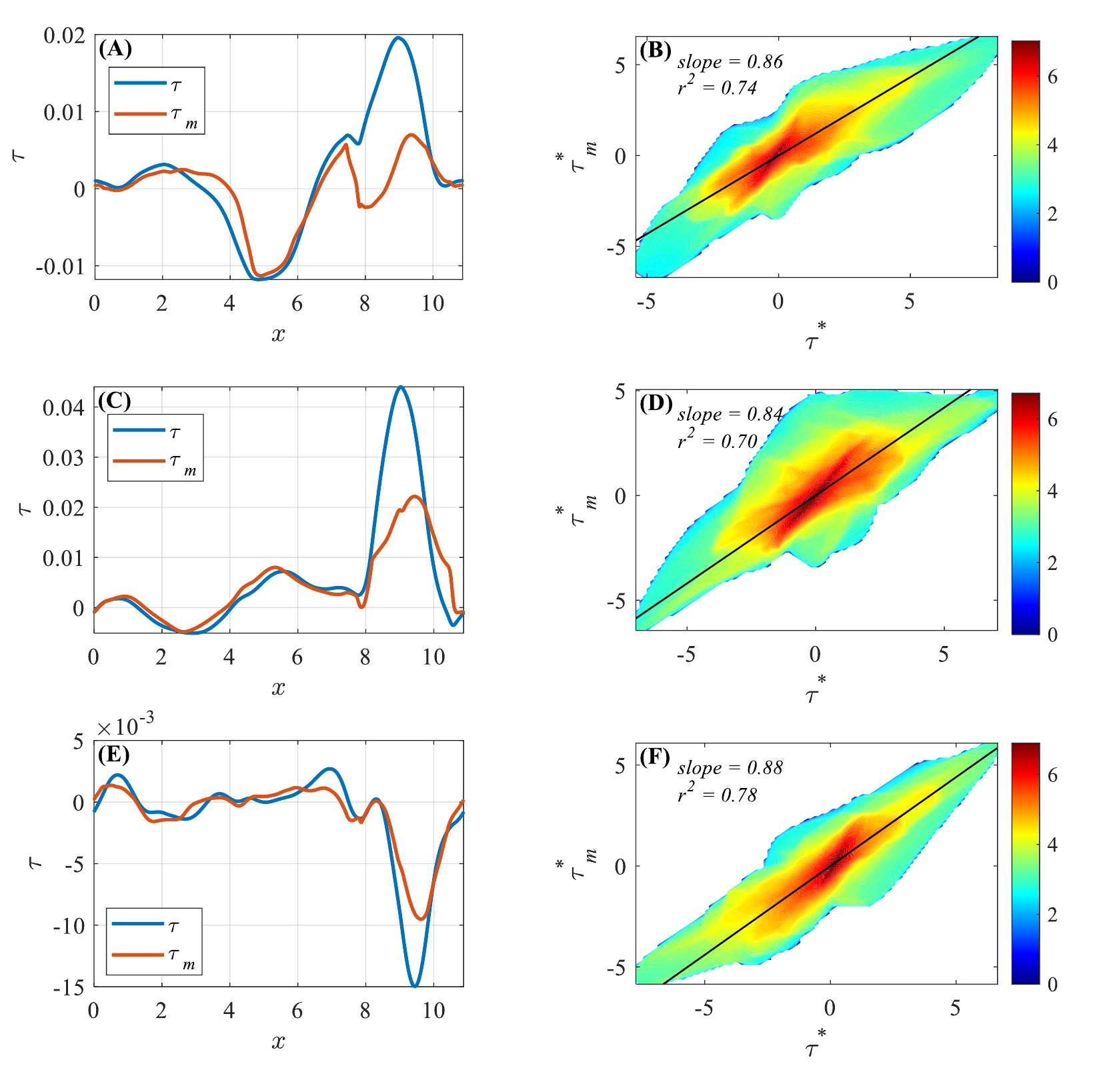}
    \caption{ Same as Fig. \ref{fig:CorrTauRhoU_k32} for $\ell = 1.5708$.}
    \label{fig:CorrTauRhoU_k4}
\end{figure}

\begin{figure}
    \centering
    \includegraphics[width = \textwidth]{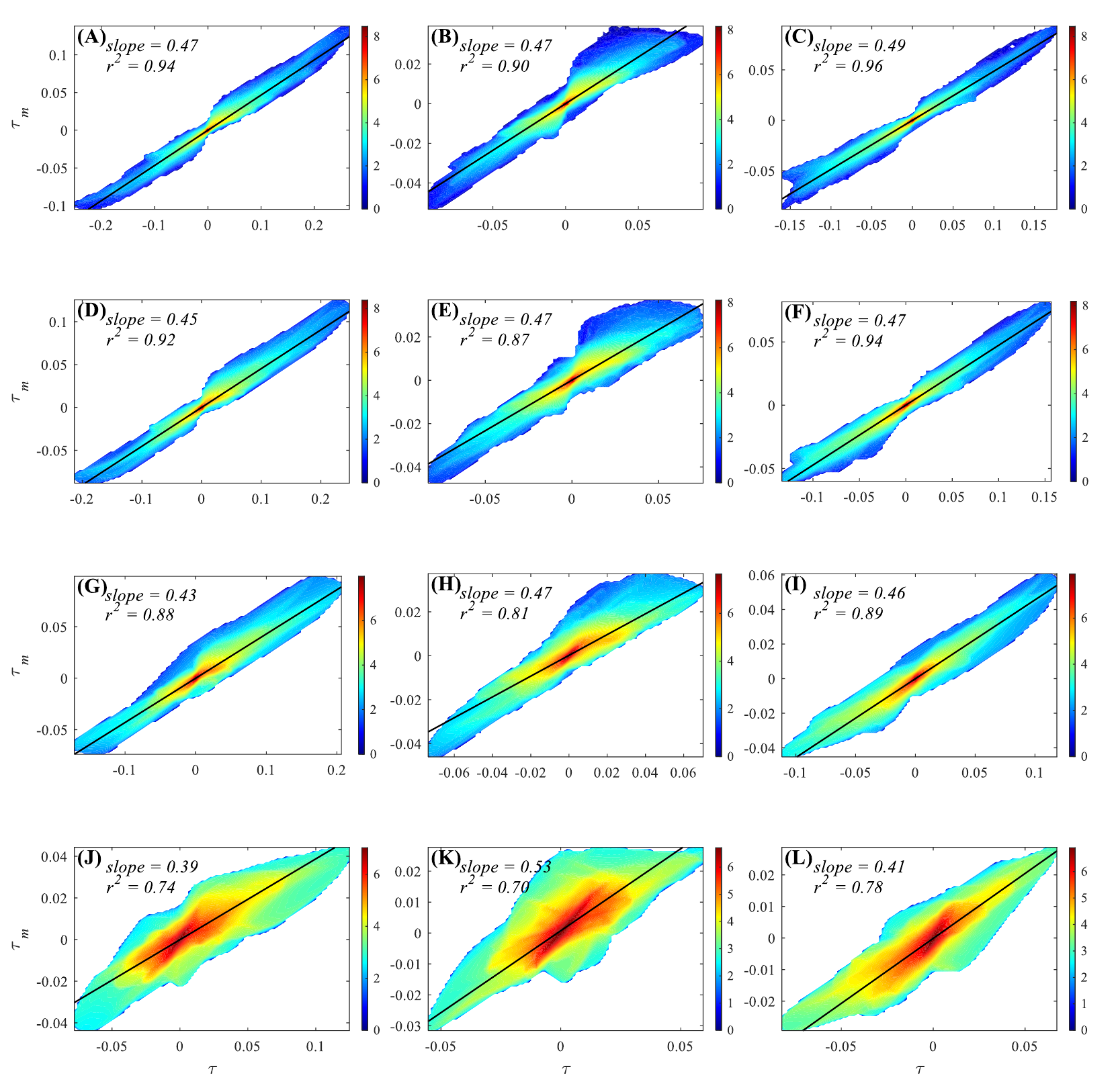}
    \caption{Similar to the joint PDFs of the z-scores in Fig.~\ref{fig:CorrTauRhoU_k32}, here we show the joint PDFs of the actual $\OL\tau_\ell(\rho, \bu)$ and its approximation $\tau_m = \frac{1}{3}M_2\ell^2 \partial_k \OL\rho \,\partial_k \OL\bu$. Top row (A, B, C) shows the joint PDFs between $\tau$ and $\tau_m$ at at $\ell=0.19635$. Second row (D, E, F) is for $\ell = 0.3927$. Third row (D, H, I) is for $\ell = 0.7854$. Bottom row (J, K, L) is for $\ell = 1.5708$. Left, middle, and right columns show, the x, y and z components, respectively.}
    \label{fig:CorrTauRhoU_k32-nostar}
\end{figure}

\begin{figure}
    \centering
    \includegraphics[width = 0.49\textwidth]{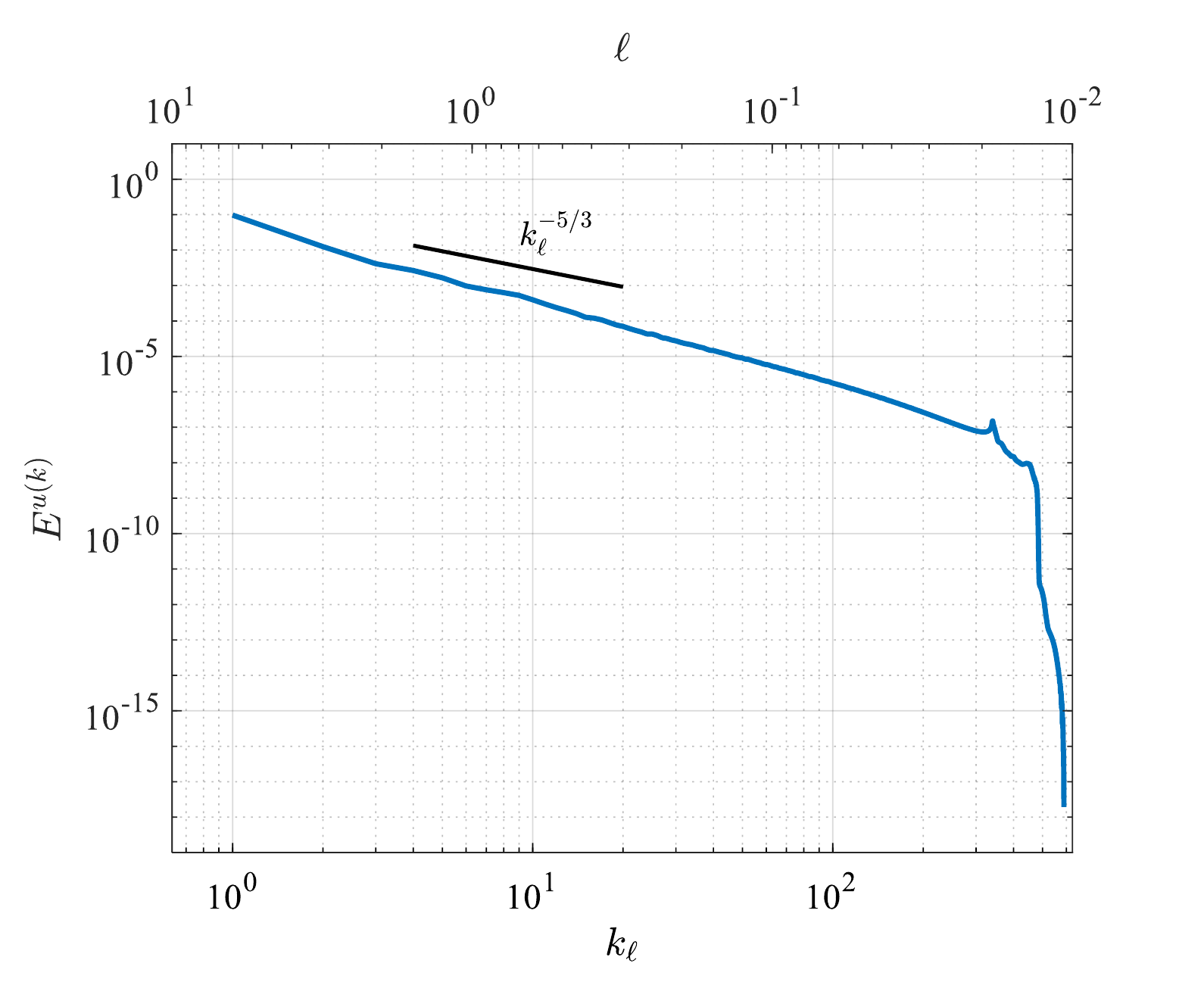}
    \includegraphics[width = 0.49\textwidth]{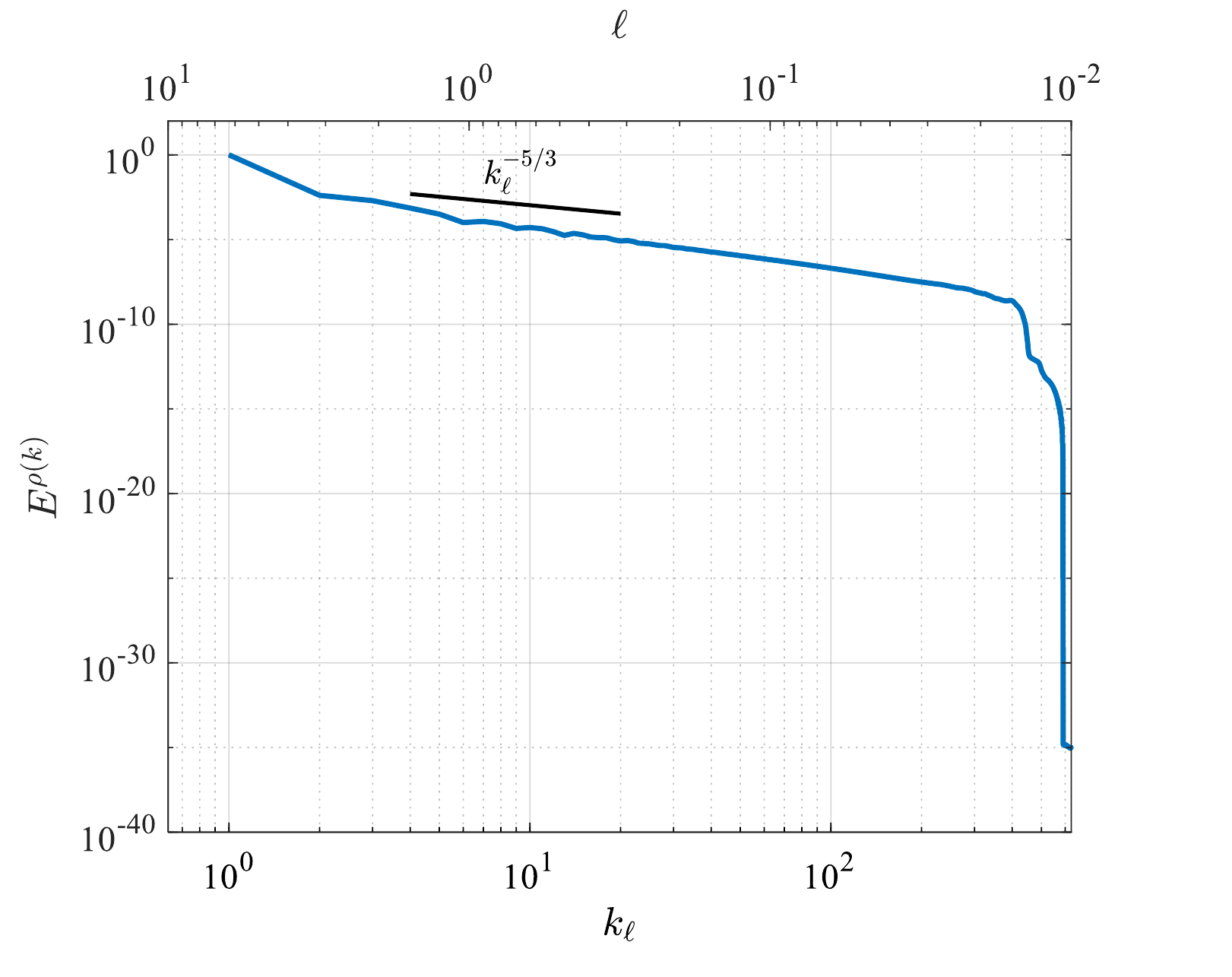}
        \caption{ Spectra of (left) velocity, $E^{u}(k)$, and (right) density, $E^{\rho}(k)$. Both decay with wavenumber $k$ sufficiently faster than the $k^{-1}$ required for the multiscale gradient expansion to converge. Here $E^{u}(k) = \sum_{\big||\bk|-k\big|<0.5}\frac{1}{2}|\hat\bu(\bk)|^2$ and $E^{\rho}(k) = \sum_{\big||\bk|-k\big|<0.5}|\hat\rho(\bk)|^2$, where $\hat\bu(\bk)$ and $\hat\rho(\bk)$ Fourier coefficients.}
    \label{fig:Spectrum}
\end{figure}

\begin{figure}
    \centering
    \includegraphics[width = \textwidth]{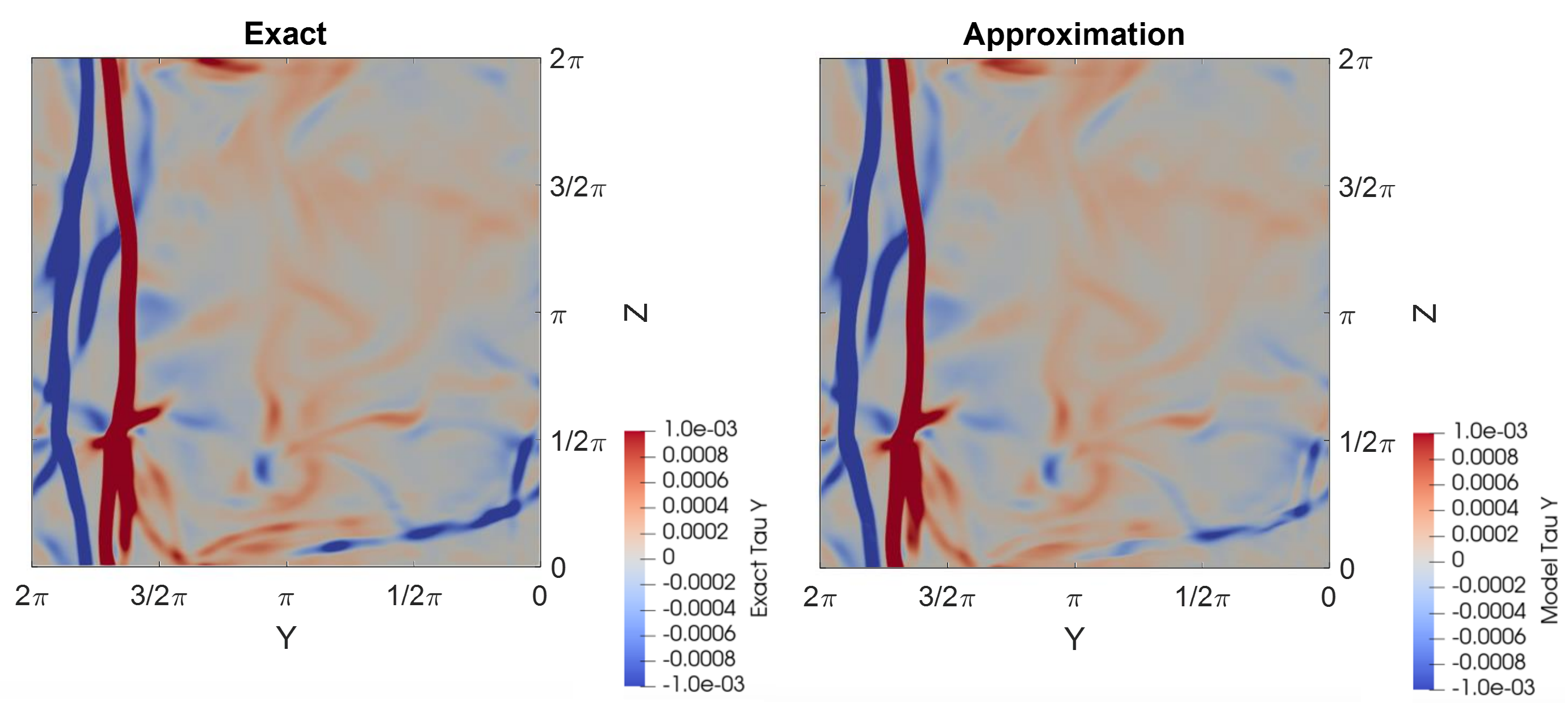}
    \includegraphics[width = \textwidth]{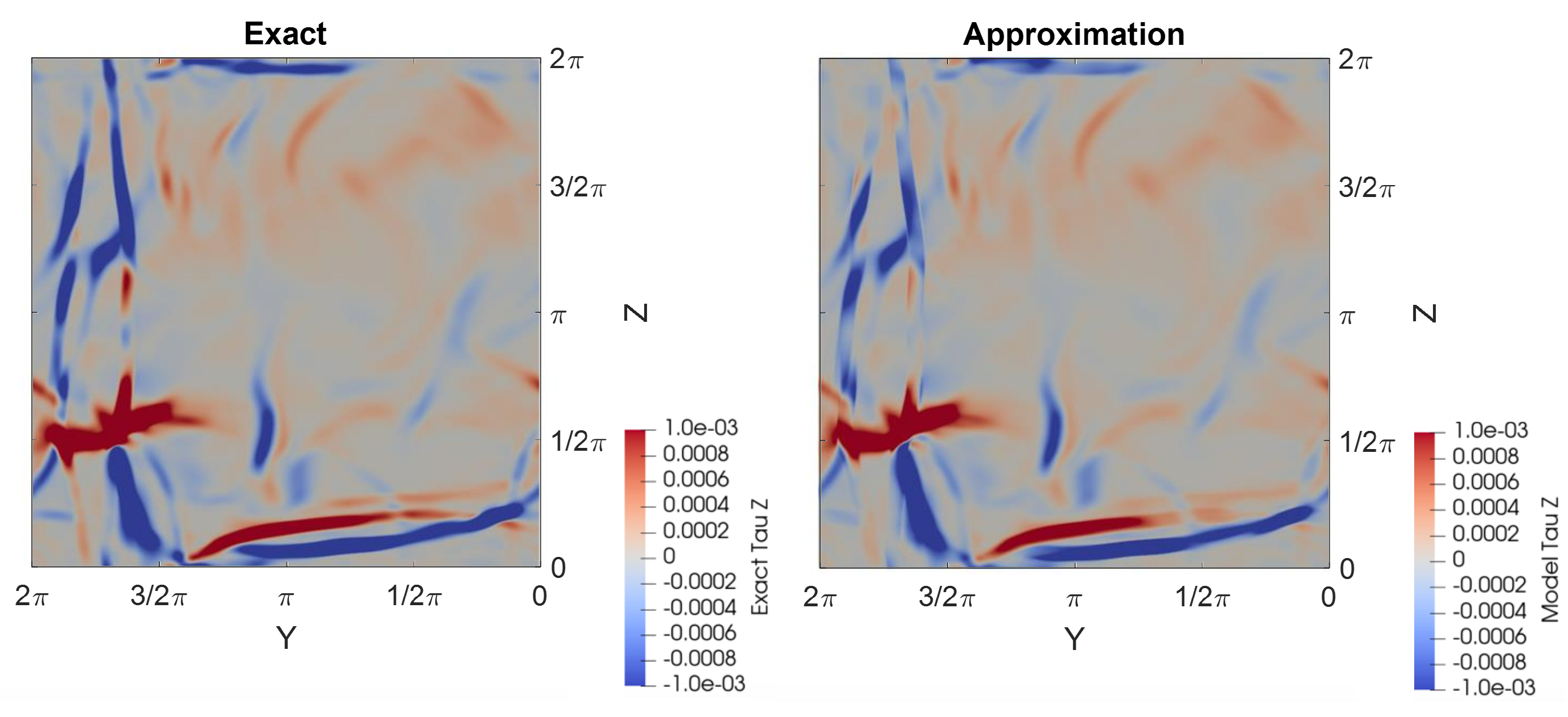}
    \caption{Same as in Fig. \ref{fig:TauVizX}, comparing the (left column) exact $\OL{\tau}_\ell(\rho,\bu)$ with (right column) its approximation for the (top row) y-components and (bottom row) z-components.}
    \label{fig:TauVizY}
\end{figure}

\begin{figure}
    \centering
    \includegraphics[width = \textwidth]{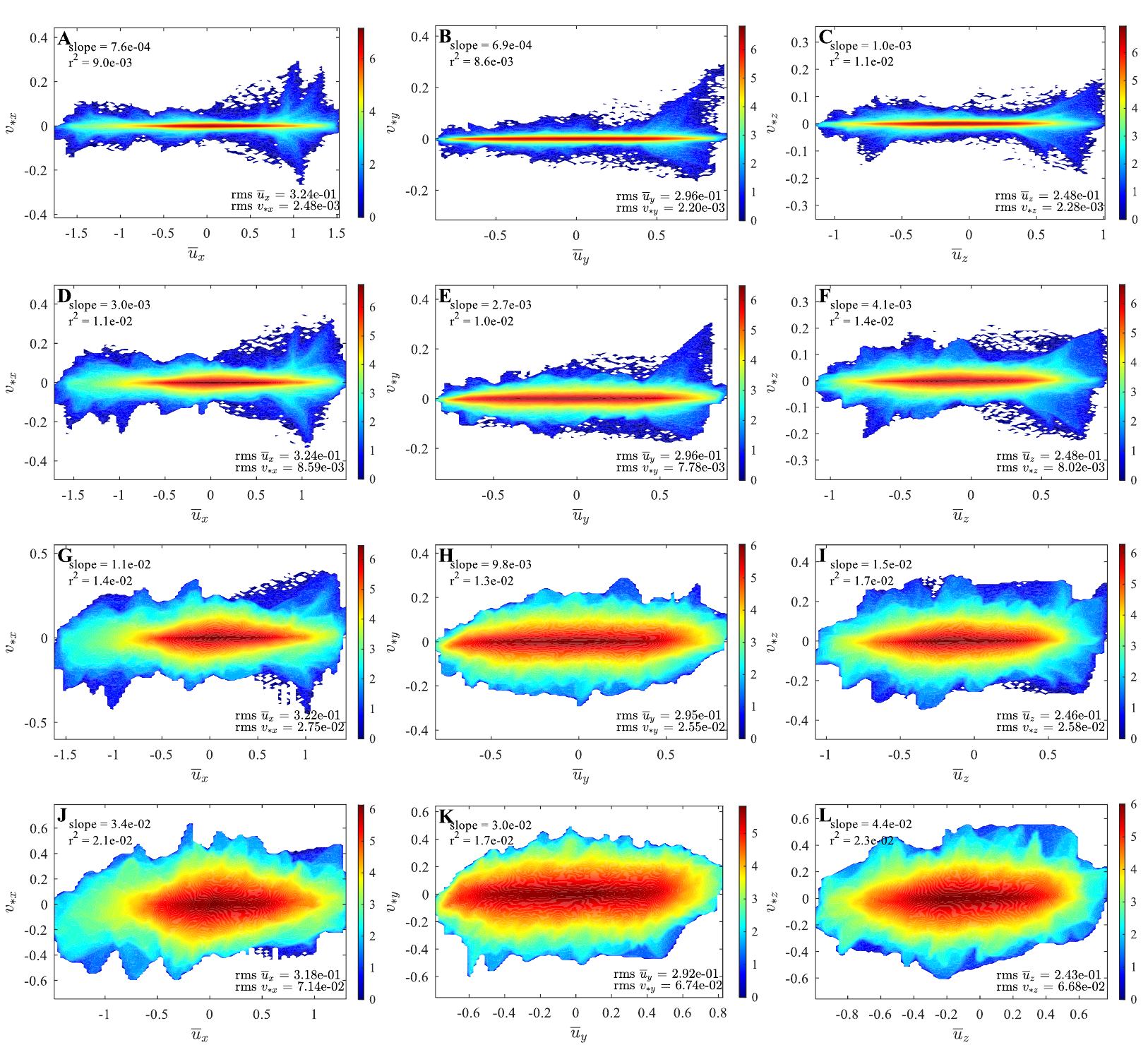}
    \caption{Joint PDFs of $\bv_*$ and $\OL \bu$. Color bar indicates the number of grid-points on a logarithmic scale, proportional to the probability. Slope of best linear fit and correlation-squared, $r^2$, are shown in the top left of each panel. The best linear fits are omitted in the the plots due to small slopes. The rms values are shown in the bottom right of each panel. Top row (A, B, C) shows the joint PDFs between $\bv_*$ and $\OL \bu$ at at $\ell=0.19635$. Second row (D, E, F) is for $\ell = 0.3927$. Third row (D, H, I) is for $\ell = 0.7854$. Bottom row (J, K, L) is for $\ell = 1.5708$. Left, middle, and right columns show, the x, y and z components, respectively. This figure highlights that the magnitudes of $\bv_*$ and the coarse velocity $\OL{\bu}_\ell$ are often comparable.}
    \label{fig:v_star_OL_u_PDF}
\end{figure}

\clearpage
\subsection{Additional supporting material from the shallow water simulation}
\begin{figure}[h]
    \centering
    \includegraphics[width = \textwidth]{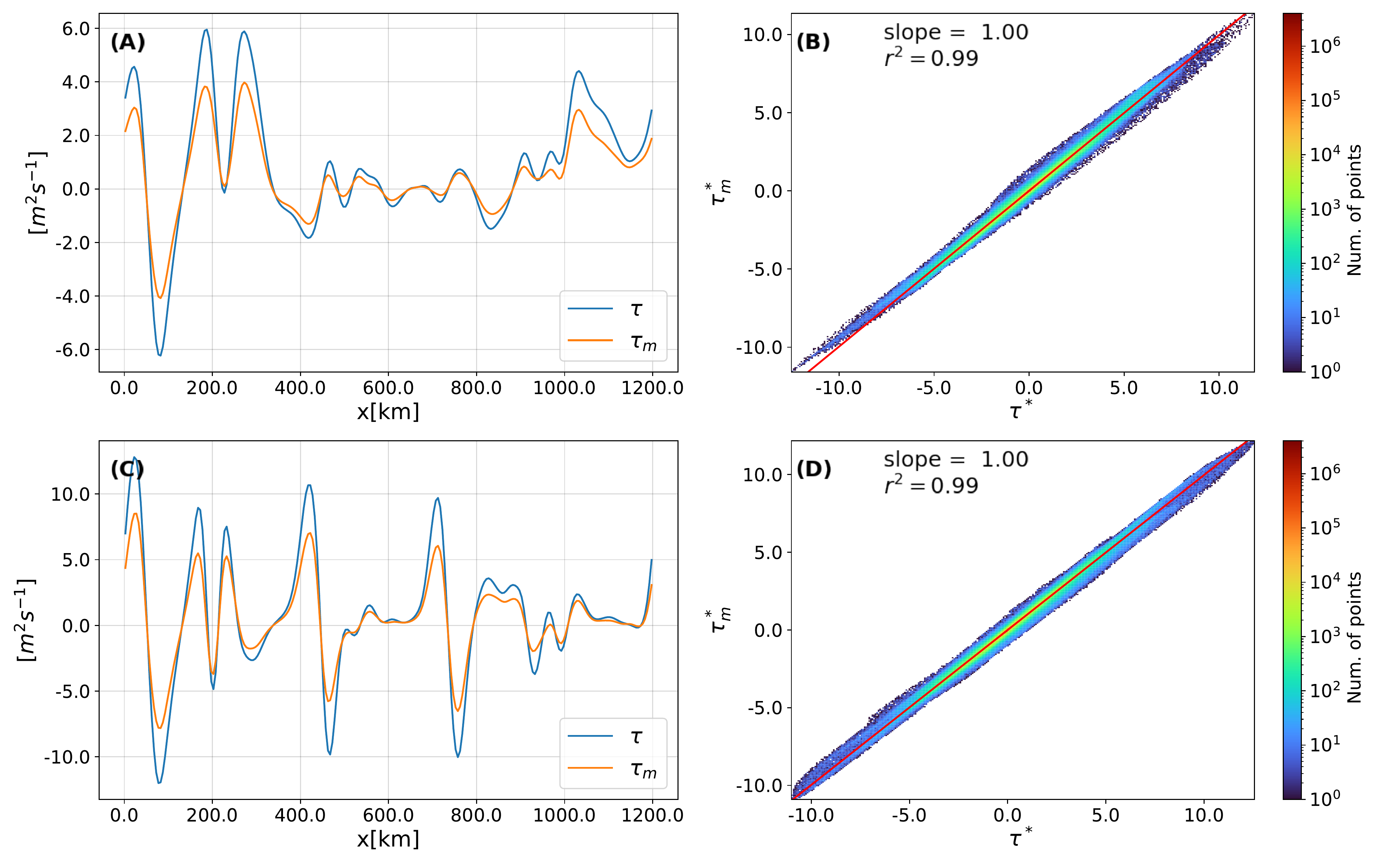}
    \caption{  Same as in Figure \ref{fig:corr100km} for $\ell = 50~km$.}
    \label{fig:corr50km}
\end{figure}

\begin{figure}
    \centering
    \includegraphics[width = \textwidth]{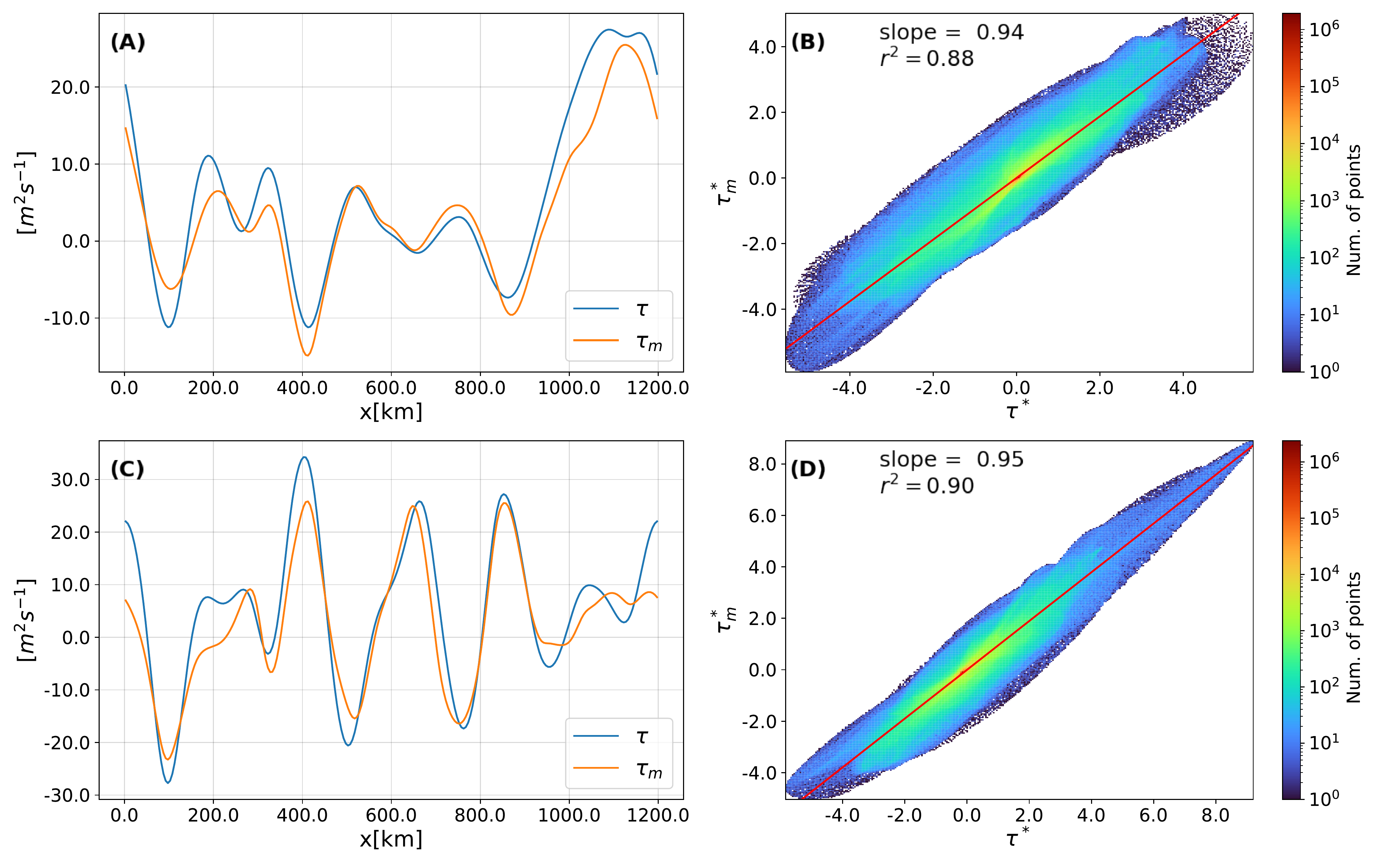}
    \caption{  Same as in Figure \ref{fig:corr100km} for $\ell = 200~km$.}
    \label{fig:corr200km}
\end{figure}

\begin{figure}
    \centering
    \includegraphics[width = \textwidth]{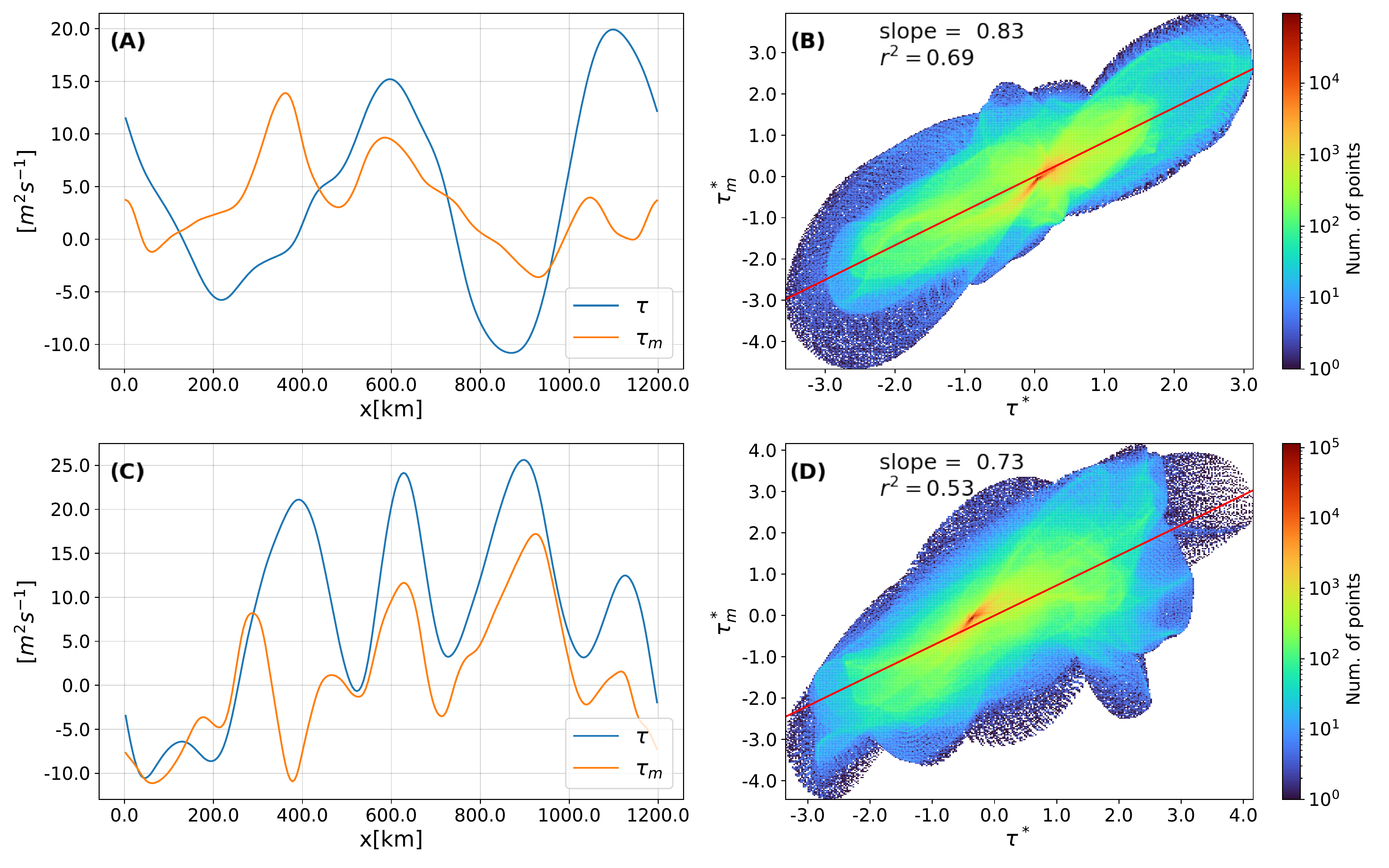}
    \caption{  Same as in Figure \ref{fig:corr100km} for $\ell = 400~km$.}
    \label{fig:corr400km}
\end{figure}

\begin{figure}
    \centering
    \includegraphics[width = \textwidth]{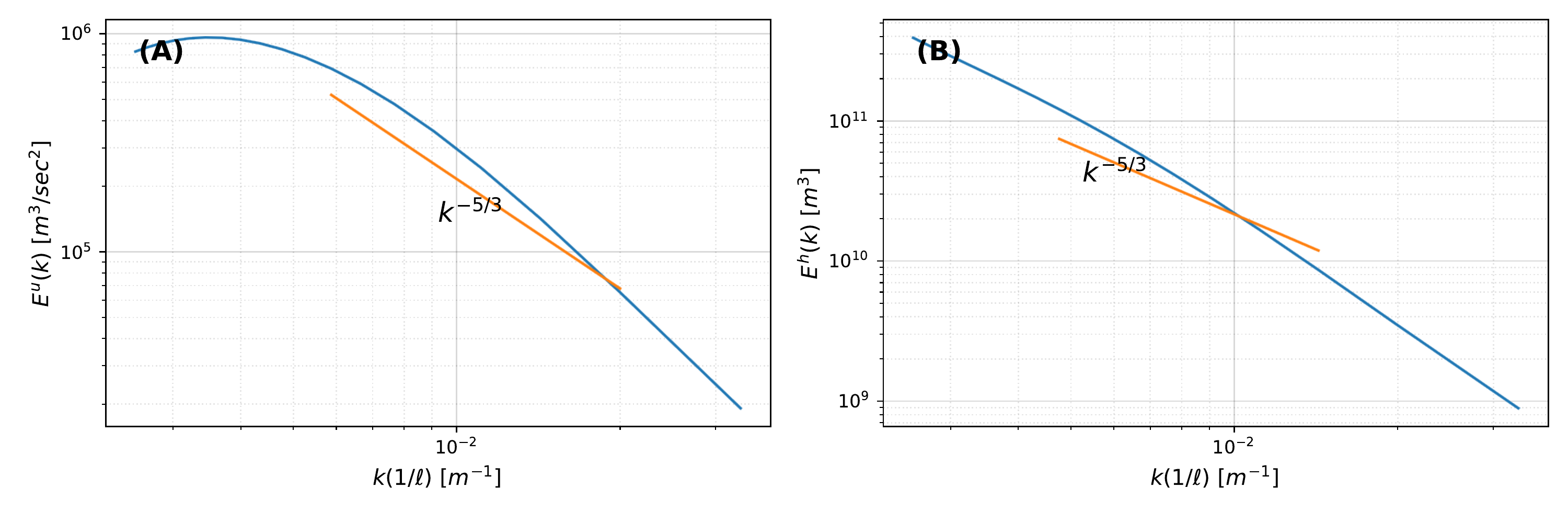}
    \caption{Spectra of (left) $\bu$ and (right) $h$ computed from the two-layer stacked shallow water simulation. These are filtering spectra, calculated by filtering at successive length-scales as described in \cite{SadekAluie18,buzzicotti2021}.}
    \label{fig:KEandh2spectra}
\end{figure}

\begin{figure}
    \centering
    \includegraphics[width = \textwidth]{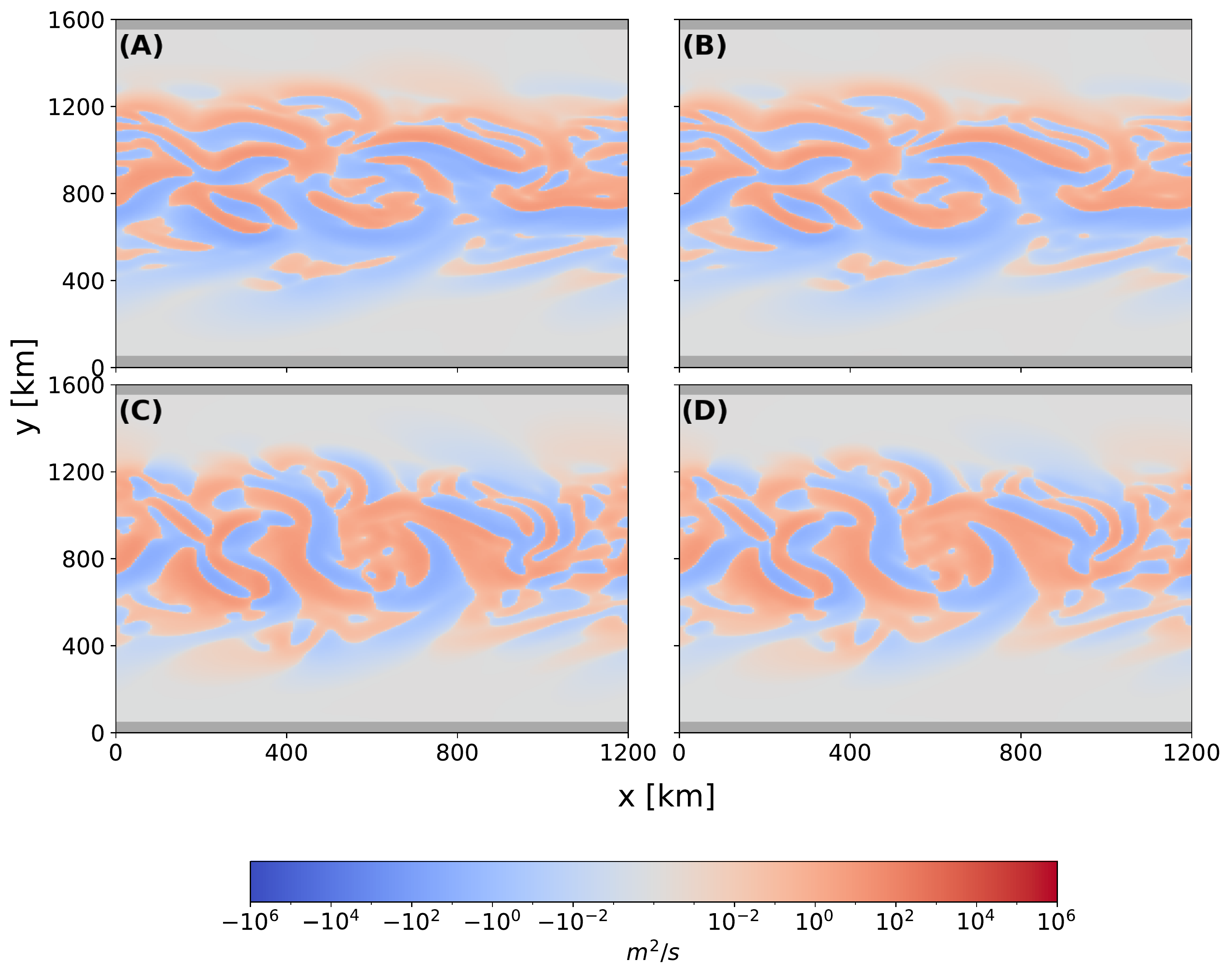}
    \caption{Same as in Fig.~\ref{fig:pmesh100km} for $\ell=50~$km.}
    \label{fig:pmesh050km}
\end{figure}

\begin{figure}
    \centering
    \includegraphics[width = \textwidth]{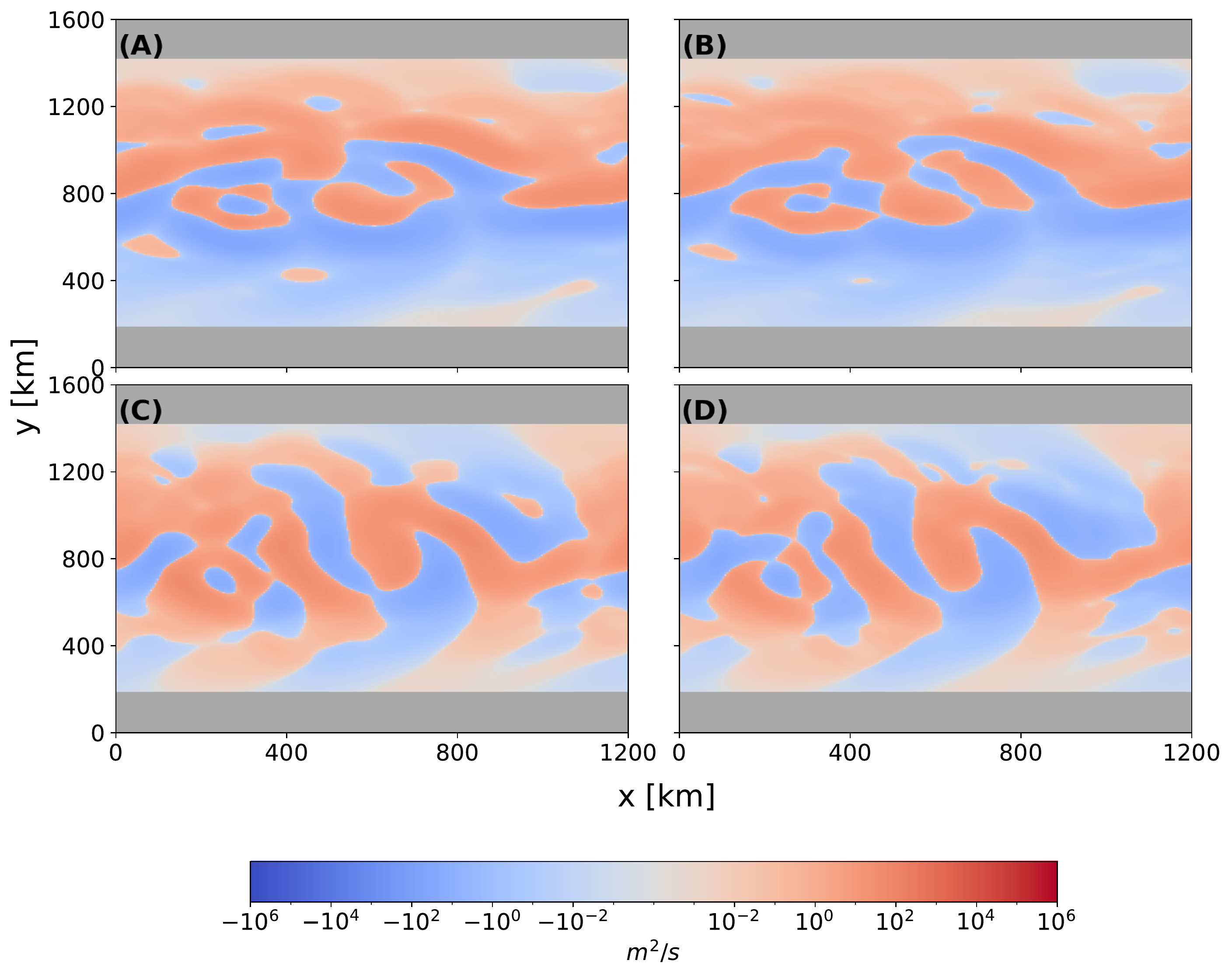}
    \caption{  Same as in Figure \ref{fig:pmesh100km} for $\ell = 200~km$.}
    \label{fig:pmesh200km}
\end{figure}

\begin{figure}
    \centering
    \includegraphics[width = \textwidth]{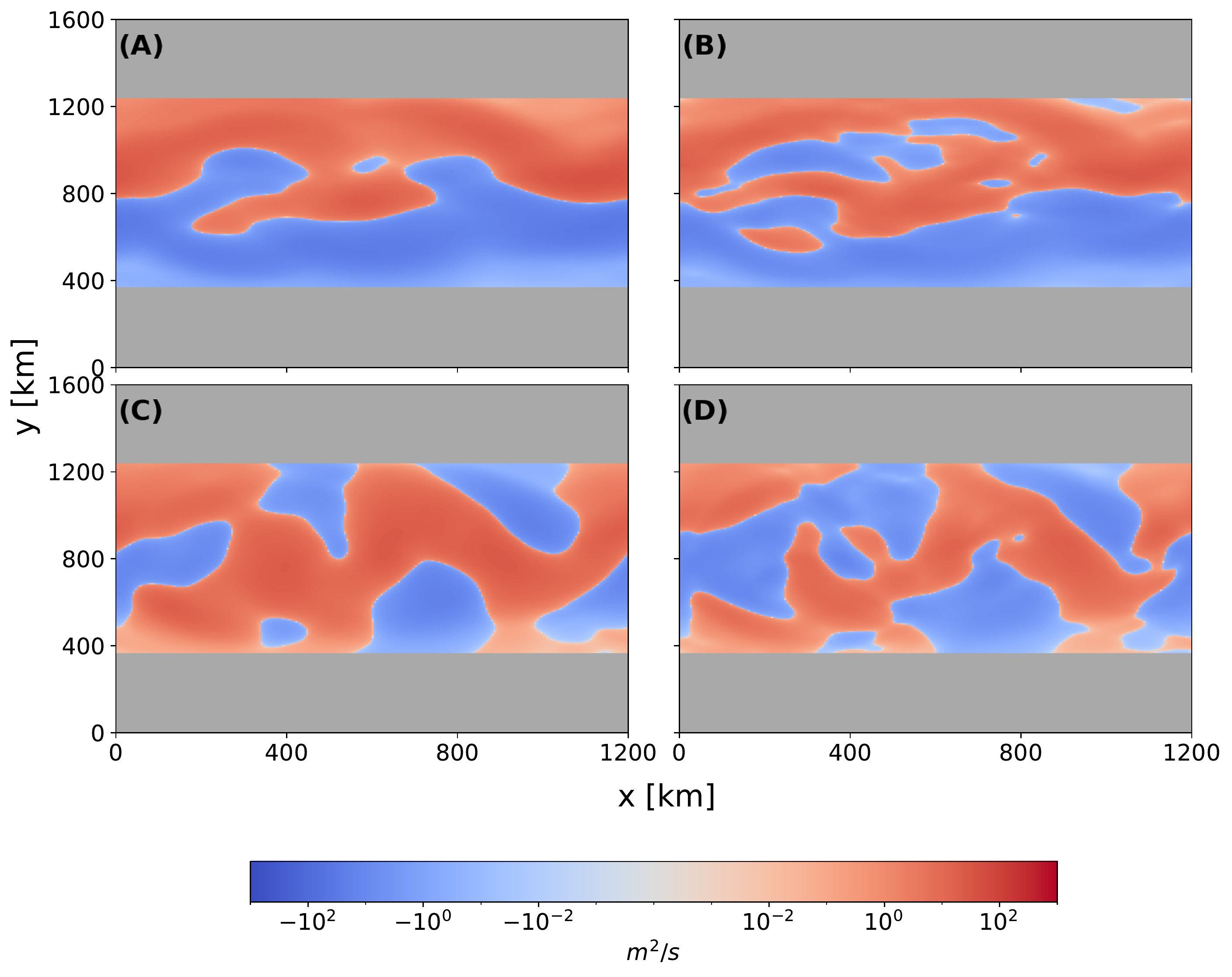}
    \caption{  Same as in Figure \ref{fig:pmesh100km} for $\ell = 400~km$.}
    \label{fig:pmesh400km}
\end{figure}

\begin{figure}
    \centering
    \includegraphics[width = \textwidth]{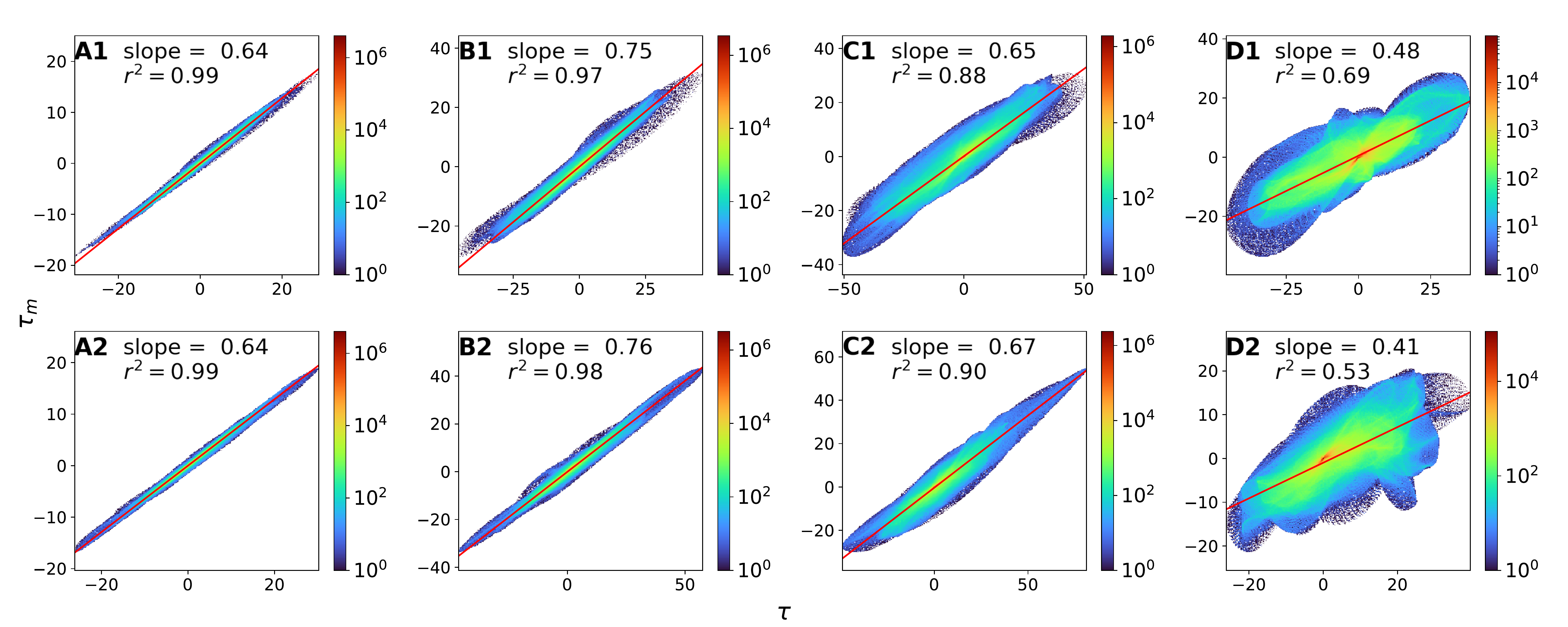}
    \caption{Similar to the joint PDFs of the z-scores in Fig.~\ref{fig:corr100km}, here we show the joint PDFs of the actual $\OL\tau_\ell(h, \bu)$ and its approximation $\tau_m = \frac{1}{2}M_2\ell^2 \partial_k \OL{h} \,\partial_k \OL\bu$, without normalization. Top and bottom panels show zonal (eastward) and meridional (northward) components, respectively. 
    A1 and A2 are for $\ell = 50~$km, B1 and B2 for $100~$km, C1 and C2 for $200~$km, D1 and D2 for $400~$km.}
    \label{fig:pdensityPlotWithoutNormalization}
\end{figure}

\begin{figure}
    \centering
    \includegraphics[width = \textwidth]{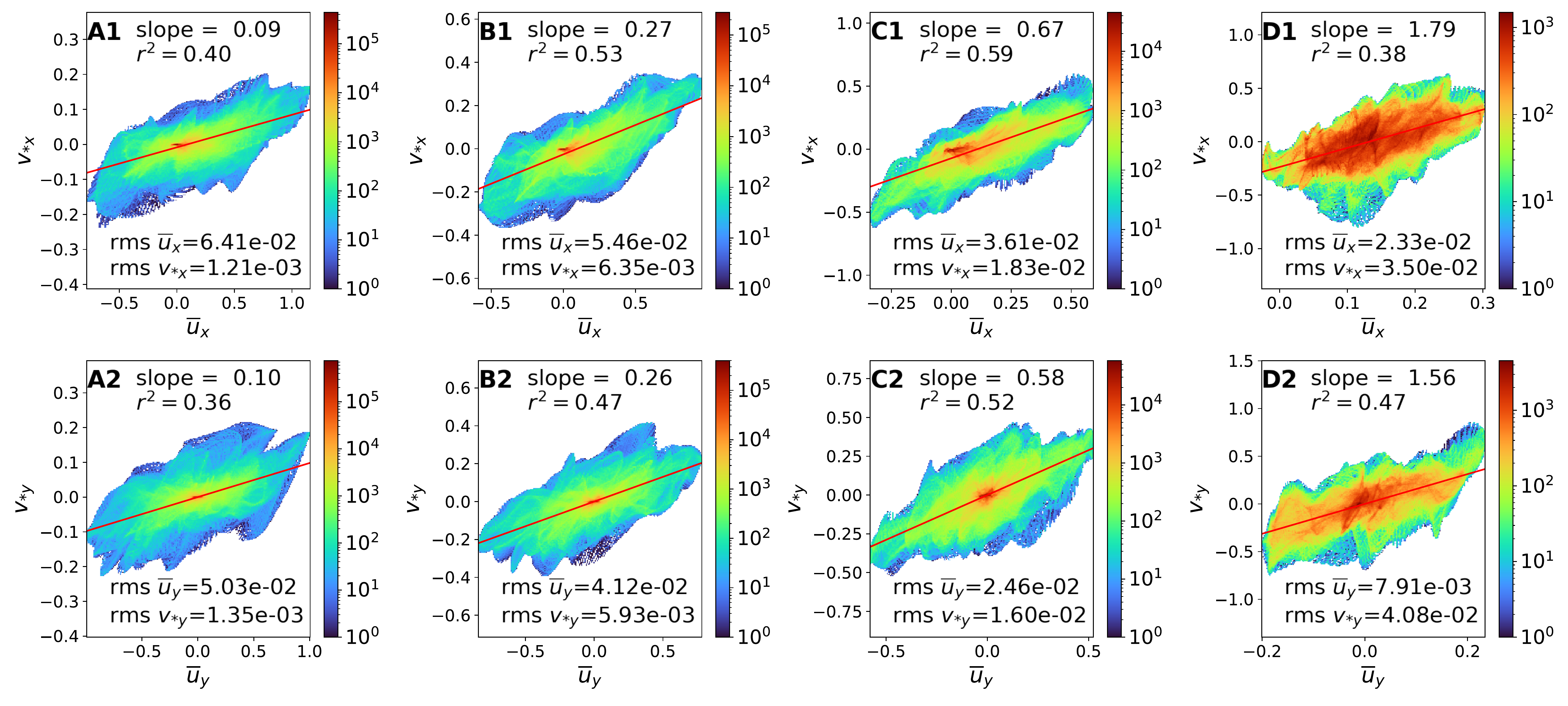}
    \caption{Joint Probability Density of $\bv_*$ and $\overline{\bu }$. The colorbars are in log scale. The top panels show zonal(eastward) componet and the bottom panels show meridional(northward) component. A1 and A2 are for $\ell$ = 50 km, B1 and B2 are for $\ell$ = 100 km, C1 and C2 are for $\ell$ = 200 km, D1 and D2 are for $\ell$ = 400 km. The red lines are linear regression fits of which the slope and goodness of fit($r^2$) are in the text in each panel. The rms values are shown in the bottom right of each panel. This figure highlights that $\bv_*$ is often comparable in magnitude relative to the coarse velocity $\OL{\bu}_\ell$.}
    \label{fig:pdfVStarVsOL_v}
\end{figure}


\end{document}